\documentclass[twocolumn,secthm,amsthm]{autart}    
\usepackage{amsfonts}
\usepackage{amssymb}
\usepackage{bbm}
\usepackage{amsthm}
\usepackage{algorithm}
\usepackage{algorithmicx}  
\usepackage{algpseudocode} 
\usepackage{graphicx}          
\usepackage{amsmath}
\usepackage{booktabs} 
\usepackage{epsfig}
 \usepackage{subfigure}
\usepackage{float}
\usepackage{stfloats}
\usepackage{pifont}
\usepackage{color}
\usepackage{etoolbox}

 \usepackage{threeparttable}

 \theoremstyle{plain}
 \newtheorem{definition}{Definition}[section]
 \newtheorem{remark}{Remark}[section]
   
 \newtheorem{theorem}{Theorem}[section]
 \newtheorem{proposition}{Proposition}[section]

\makeatletter

\g@addto@macro\normalsize{%
  \setlength{\abovedisplayskip}{5pt plus 1pt minus 1pt}%
  \setlength{\belowdisplayskip}{5pt plus 1pt minus 1pt}%
  \setlength{\abovedisplayshortskip}{3pt plus 1pt minus 1pt}%
  \setlength{\belowdisplayshortskip}{3pt plus 1pt minus 1pt}%
}

\normalsize

\makeatother

\makeatletter
\renewenvironment{proof}[1][\proofname]{%
  \par\vspace{2pt plus 0.5pt minus 0.5pt}%
  \pushQED{\qed}%
  \normalfont
  \noindent\textit{#1\@addpunct{.}}\hspace{0.5em}%
  \ignorespaces
}{%
  \popQED
  \par\vspace{2pt plus 0.5pt minus 0.5pt}%
}
\makeatother

\begin{document}

\begin{frontmatter}

\title{A Class of Bayesian Conjugate Distributions on $SE_n(3)$ for Motion Estimation \thanksref{footnoteinfo}} 

\thanks[footnoteinfo]{This paper was not presented at any IFAC 
meeting. This work was supported in part by the National Natural Science
Foundation of China under Grant 12422214. Tel. +86-10-61716698, Fax. +86-10-61716878}

\author[Shenyuan]{Shijie Wang}\ead{wangsj2001@buaa.edu.cn},    
\author[Astronautics_Gui]{Haichao Gui}\ead{hcgui@buaa.edu.cn},               
\author[Astronautics_Gui]{Rui Zhong}\ead{zhongruia@163.com}  

\address[Shenyuan]{School of Astronautics, Shenyuan Honors College, Beihang University, Beijing, 100191, China}
\address[Astronautics_Gui]{School of Astronautics and the MOE Key Laboratory of Spacecraft Design, Optimization and Dynamic Simulation, Beihang University, Beijing, 100191, China}


\begin{abstract}                          
This paper proposes a novel class of Bayesian Conjugate Distributions (BCDs) on extended special Euclidean groups $SE_n(3)$ for global motion uncertainty representation and estimation. BCDs are constructed by restricting matrix Gaussian distributions on ambient Euclidean space to $SE_n(3)$, avoiding singularity of local coordinates. The unique feature of BCDs is their conjugacy to a broad class of sampling distributions arising in motion estimation, enabling exact closed-form Bayesian inference. Optimal estimation with respect to BCDs is further investigated for implementing motion estimation. Two means are extracted as random state estimates, which maximize probability and minimize mean squared error, respectively. Moreover, within the BCD class, moment matching is shown to minimize the Kullback–Leibler divergence, and maximum likelihood estimation is also derived, which provide analytical methods to approximate underlying distributions with the optimality guarantee. As an application, a Bayesian estimator for pose and velocity estimation of an Unmanned Aerial Vehicle (UAV) is developed within an assumed density filtering architecture. The propagation and estimate extraction leverage the optimal estimation related to BCDs, and the landmark measurements are exactly fused with the BCD prior in closed form due to the conjugacy. Numerical simulations of a UAV flight indicate the advantages of BCD-based estimators over two classic Kalman filters on Lie groups with various initial uncertainties and measurement noise.
\end{abstract}

\end{frontmatter}

\section{Introduction}
\label{sec:introduction}
Motion estimation based on sensor fusion plays a crucial role in many engineering fields, including robotics\cite{burnett2024continuous}, aerospace\cite{whittaker2025inertial}, etc. Since motion states involving attitude evolve on nonlinear manifolds with inherent constraints, conventional nonlinear filtering or optimization methods may fail to achieve accurate estimation if the underlying geometric structure is not properly considered. Meanwhile, sensors are inevitably affected by random noise, whose effects cannot be neglected and need to be carefully modeled.


When incorporating the kinematics and dynamics of systems, the main branches of motion state estimators are deterministic nonlinear observers and Kalman filters on Lie groups. Deterministic observers are directly designed on state manifolds, avoiding the singularity of local coordinates and achieving global or almost global stability for attitude estimation, pose estimation, and more complex situations \cite{berkane2017hybridbias,wang2020hybrid,wang2023nonlinear}. While these estimators are distinguished by their high reliability and a sound theoretical framework, this paper focuses on estimation techniques accounting for the stochastic properties of states. Among stochastic motion estimators, an important development was the multiplicative extended Kalman filter (MEKF)\cite{leffertsKalmanFilteringSpacecraft1982,markley2003attitude}, which represents attitude by unit quaternions and approximates uncertainty in Euclidean space by Gaussian distributions. The invariant extended Kalman filters (IEKFs) further introduce the Lie group structure of the motion state space, as developed in \cite{barrau2016invariant,barrauInvariantKalmanFiltering2018,barrau2022geometry}. IEKFs use intrinsic properties of Lie groups to linearize nonlinear systems and characterize the uncertainty of the state via a Gaussian distribution in exponential coordinates. IEKFs can provide estimates with lower errors and faster convergence rates than those of the MEKF, benefiting from their better consistency with geometric and dynamic structures. Similar methodologies are also applied to the Unscented Kalman filter (UKF) \cite{brossard2017unscented} and Particle Filter (PF) \cite{jin2024computationally}.

Concentrated Gaussian Distributions (CGDs), as presented in \cite{wang2006error,chirikjian2011stochastic}, are defined on the Lie group via formulating Gaussian distributions on the Lie algebra with the exponential map. CGDs are used to construct Kalman filters on Lie groups \cite{bourmaud2015continuous} and also to characterize the uncertainty of states in the IEKF \cite{barrauInvariantKalmanFiltering2018}. Although CGDs are described as distributions on Lie groups, the fusion and propagation involved with CGDs generally require approximation based on the Baker--Campbell--Hausdorff (BCH) formula and local coordinate representations \cite{barfoot2014associating,bourmaud2015continuous}. Moreover, while pose graph optimization in Simultaneous Localization and Mapping (SLAM) and robotics is designed towards global optimality, its front-end preintegration and back-end optimizer often rely on local representations of uncertainty \cite{brossard2021associating,barfoot2025certifiably}.

Recently, there are several attempts to develop global probabilistic representations on the state manifold, and to apply them to Bayesian estimation. Matrix Fisher distributions  (MFDs) \cite{leeBayesianAttitudeEstimation2018} and Bingham distributions \cite{gilitschenski2015unscented} are two primary choices for pure attitude estimation. Additionally, their joint distributions with Gaussian distributions, in the form of conditional–marginal distributions, are used to simultaneously estimate the attitude and the states in the n-dimensional Euclidean space $\mathbbm{R}^n$, such as sensor bias \cite{wang2021matrix,darling2016uncertainty}. Despite the fact that the three-dimensional special Euclidean group $SE(3)$ and its extension $SE_n(3)$ play important roles in motion state representation, distributions defined directly on these groups remain limited, such as \cite{oualkacha2012estimation} and \cite{ley2018applied}. These distributions on $SE(3)$ miss coupling terms between the attitude and the translation component, and have not yet been used for Bayesian pose estimation. In designing estimators with these distributions, propagation along the dynamic or kinematic equations of the system, fusion with measurement distributions, and the extraction of the optimal state estimates constitute three fundamental steps. The propagation of distributions is related to obtain the optimal approximation of underlying distributions in a certain distribution class, while the fusion involves the conjugacy \cite{gelman1995bayesian}, allowing the posterior distribution to remain within the same class of the prior and thus enabling tractable recursive estimation.  For Bayesian estimators with MFDs \cite{leeBayesianAttitudeEstimation2018}, the MFD approximation of propagated distributions computed by moment matching was proven optimal in terms of maximum likelihood estimation, and they are also shown to be conjugate to von Mises Fisher and Gaussian distributions in \cite{leeBayesianAttitudeEstimation2018,wang2025attitude}, which are used to characterize vector measurements. While matrix Fisher-Gaussian distributions (MFGDs) proposed in \cite{wang2021matrix} carefully design the correlation terms between the attitude and states in the Euclidean space with an intuitive geometric interpretation, MFGD-based estimators cannot provide globally optimal approximation of propagated distributions and no distribution, to the best of our knowledge, have been found to be conjuagated to MFGDs, including themselves. The Bingham-Gaussian distributions (BGDs) in \cite{wang2021bingham} and \cite{darling2016uncertainty} suffer from similar issues.

Due to the absence of conjugacy to common probabilistic models of motion measurements, the approximation of the underlying posterior distributions needs to be explicitly considered in the Bayesian estimator with these joint distributions. For certain specific scenarios, such as attitude-bias estimation with unit vector measurements in \cite{wang2021matrix} or pose estimation with mutually independent position and attitude measurements \cite{wang2023ins}, tailored approximation strategies can be derived to obtain tractable posterior updates. These strategies are inherently problem-dependent and inevitably result in a loss of theoretical accuracy. When more general measurement models are involved, especially those containing position-attitude coupling terms such as landmark observations, constructing accurate approximations of the underlying posterior distributions with joint distributions defined on state manifolds, such as MFGDs and BGDs, is still an open problem. Moreover, obtaining an analytical solution for optimal parameter estimation, including maximum likelihood estimation or minimum Kullback--Leibler (KL) divergence estimation of these distributions remains insufficiently addressed in existing literature, despite being essential for Bayesian estimators under the assumed density framework. Existing approaches either rely on marginal–conditional approximations for MFGDs \cite{wang2021matrix} or require numerical optimization on manifolds for BGDs \cite{darling2016uncertainty}. 

To address the absence of conjugacy and analytical solutions for optimal parameter estimation, this paper develops a conjugate probabilistic distribution class for Bayesian motion estimation on $SE_n(3)$, which is conjugate for common probabilistic models of motion measurements and enables analytical and optimal parameter estimation from moments or samples. Based on this distribution class, we further design a Bayesian pose estimator that incorporates landmark measurements, to demonstrate the applicability of the proposed distributions to measurements involving position--attitude coupling. The three main contributions are summarized as follows.

First, this paper proposes a novel class of Bayesian Conjugate Distributions (BCDs) in Section \ref{sec:def}, which globally represent the uncertainty of motion states on $SE_n(3)$. The proposed distribution class is defined by confining Gaussian distributions on the ambient space of $SE_n(3)$, and is designed to be conjugate to itself and a broad range of distributions characterizing measurements used in regular pose estimation tasks. The introduction of conjugacy enables rigorous and precise closed-form Bayesian inference in the form of the BCD class when fusing both pure position-/attitude-related measurements and position-attitude-coupled measurements, including GPS, magnetometer, and known and unknown landmark measurements.

In Section \ref{sec:opt_est}, the optimal estimation of state and parameters is developed under the BCD representation. Two estimates are given for states characterized by BCDs, which are derived from the maximum probability and minimum squared error optimizations, respectively. Then, we consider obtaining the globally optimal BCD approximation of the underlying distributions using moments and samples. All parameters of a BCD are shown to be obtained from finite-order moment matching, and the distribution with these parameters is proven to be the minimum KL divergence estimation of the underlying distribution. The joint Maximum Likelihood Estimation (MLE) with samples on $SE_n(3)$ is then presented, avoiding the introduction of local sampling from Gaussian distributions on $\mathbbm{R}^n$ in \cite{jin2024computationally} or conditional-marginal MLE as an approximation in \cite{wang2021matrix}. The solutions of both moment matching and MLE are derived analytically. Most parameters are computed in closed form, while the rest are solved from equations in $\mathbbm{R}^3$, thereby avoiding the numerical optimization on manifolds, as in \cite{darling2016uncertainty}.

A BCD-based Bayesian estimator for estimating the pose and linear velocity of a UAV with known landmarks is presented, demonstrating the applicability of the proposed distributions to motion estimation problems with position-attitude coupling measurements. The BCD-based estimator follows the structure of assumed density filtering and includes propagation, fusion, and estimate extraction steps, illustrating how the properties in Section \ref{sec:def} and \ref{sec:opt_est} are utilized in estimation algorithm construction. In the proposed estimator, the propagation step provides the minimum KL-divergence BCD prior approximation based on propagated moments, and the fusion with sampling distributions from measurement models is analytical and structure-preserving due to the conjugacy. The maximum probability and minimum squared error estimates are used as extracted estimates, respectively. In the simulation of a UAV flight presented in Section \ref{sec:Sim}, we observe that the BCD-based estimator provides lower RMSEs and better transient response than those of two classical Kalman filters on $SE_n(3)$. Moreover, the estimator with minimum squared error estimates demonstrates slightly better performance compared to the estimator with BCDs using maximum probability estimates in the mean RMSE over 50 Monte Carlo simulations, indicating its long-term accuracy.

\section{Preliminaries}
\subsection{Notations and Basic Facts}
Denote by $\mathbbm{R}^n$ the n-dimensional Euclidean space, $\mathbbm{N}^+$ the set of positive integers, and $I_n$ the $n\times n$ identity matrix. Let $\mathcal{F}_I$ be an inertial frame with basis $\mathcal{F}_I=\{e_1,e_2,e_3\}$, and $\mathcal{F}_B$ be a frame attached to a rigid body moving in three-dimensional space. The definition for the special orthogonal group reads $ SO(3):=\left\{   R\in\mathbbm{R}^{3\times 3}|RR^T=I_3,{\rm det}(R)=1\right\}$. In this paper, the extended special Euclidean group is used to represent motion states, defined as
\begin{align*}
     & SE_n(3): = \nonumber\\
      &\left\{ {\mathcal{T}(R,X) = \left[ {\begin{array}{*{20}{c}}
R&X\\
{{0_{n\times 3}}}&I_n
\end{array}} \right] \bigg | R\in SO(3),X\in \mathbbm{R}^{3\times n} } \right\},
\end{align*}
of which the group composition boils down to matrix multiplication, and the element of $SE_n(3)$ is referred to as extended transformation matrix. The set of n-dimensional symmetric matrices is denoted by $\mathbbm{S}^n$, and its positive definite and semi-definite subsets are denoted by $\mathbbm{S}^n_{++}$ and $\mathbbm{S}^n_{+}$, respectively. Similarly, its negative definite and semi-definite subsets are denoted by $\mathbbm{S}^n_{--}$ and $\mathbbm{S}^n_{-}$.  Define the vector operator for given $X=[x_1,\dots,x_n]\in \mathbbm{R}^{3\times n}$ as $\overrightarrow{X}=[x_1^T,\dots,x_n^T]^T\in \mathbbm{R}^{3n}$. The inverse operator is denoted by $\overleftarrow{x}$ for $x\in \mathbbm{R}^{3n}$.
The operator $\wedge$ turns $\xi \in \mathbbm{R}^{3+3n}$ into a member of the Lie algebra $\mathfrak{se}_n(3)$ according to
\begin{equation}
  {\xi ^ \wedge }: = {\left[ {\begin{array}{*{20}{c}}
\phi \\
\rho 
\end{array}} \right]^ \wedge } = \left[ {\begin{array}{*{20}{c}}
{{\phi_\times }}&  \overleftarrow{\rho}\\
{{0_{n \times 3}}}&0_{n\times n}
\end{array}} \right],
\end{equation}
where $\phi\in \mathbbm{R}^3$ and $\rho \in \mathbbm{R}^{3n}$, and the subscript $\times$ is defined such that $x_\times\ y = x\times y$ for any $x,y\in \mathbbm{R}^3$. The inverse of $\wedge$ is denoted by $\vee$. The vector $\xi \in \mathbbm{R}^{3+3n}$ is mapped to an extended transformation matrix $T\in SE_n(3)$ by the matrix exponential $ T(\xi)=\exp \left( \xi^{\wedge} \right)$.

The trace of a matrix is denoted by $\text{tr}(\cdot)$, and the function $e^{\text{tr}(\cdot)}$ is abbreviated as $\text{etr}(\cdot)$. The Frobenius norm of a matrix $M\in \mathbbm{R}^{n\times n}$ is defined as $\Vert M \Vert = \sqrt{\text{tr}\left( M^TM \right)}$, and the weighted F-norm distance for $A,B\in \ \mathbbm{R}^{n\times n}$ is defined as $\|A-B\|_W = \sqrt{(A-B)W(A-B)^T},$ where $W\in\mathbbm{S}^n_+$ is a symmetric positive definite matrix.

\subsection{Matrix Fisher distributions on $SO(3)$}
The matrix Fisher distribution (MFD), as a compact exponential density on the Stiefel manifold,  has been applied to represent uncertainty on $SO(3)$ for attitude estimation \cite{leeBayesianAttitudeEstimation2018}. Although the proposed distributions are not constructed directly based on MFDs, the following properties are useful for subsequent derivations. The distribution of a random rotation $R$ following a MFD with parameter $F\in \mathbbm{R}^3$, denoted by $R\sim\mathcal{M}(F)$, is defined by the following Probability Density Function (PDF)
\begin{align}
     p(R\in SO(3);F) = \frac{1}
{{c_{\mathcal{M}}\left( F \right)}}\text{etr}\left[ {F^T R} \right],\label{eq:de_MF}
\end{align}
where $c_{\mathcal{M}}(F)$ is the normalized constant, calculated by $c_{\mathcal{M}}(F) = \int_{SO(3)} {\text{etr}\left[{F^TR}\right]dR}$ with the normalized bi-invariant Haar measure $dR$. Given the proper singular value decomposition (pSVD) \cite{markleyAttitudeDeterminationUsing1988} of $F$ as $ {\rm pSVD}(F)= (U,S,V)\in SO(3)\times \mathbbm{R}^{3\times 3}\times SO(3)$ such that $F=USV^T$, the mean attitude of $R$ is given by $M=UV^T\in SO(3)$, which maximizes the density function and minimizes the Frobenius mean squared error. Meanwhile, there is $c_{\mathcal{M}}(F)\equiv c_{\mathcal{M}}(S)$.

According to Theorem 2.2 and Lemma 2.2 in \cite{leeBayesianAttitudeEstimation2018}, the first moment, i.e., the arithmetic mean, of $R$ is given by
\begin{align}
   E(R) = U{\rm{diag}}\left[ {\frac{{\partial \ln c_{\mathcal M}(S)}}{{\partial {s_1}}},\frac{{\partial \ln c_{\mathcal M}(S)}}{{\partial {s_2}}},\frac{{\partial \ln c_{\mathcal M}(S)}}{{\partial {s_3}}}} \right]{V^T}.\label{eq:MF_E}
\end{align}
Both $c(s)$ and $\partial c(S)/\partial s_i$ can be calculated using the integral of the modified Bessel function of the first kind \cite{abramowitz1966handbook}.

\section{Bayesian Conjugate Distributions on $SE_n(3)$} \label{sec:def}

\subsection{Definitions}
The definitions of BCDs are motivated by the relationship between the MFD and the matrix Gaussian distribution presented in \cite{downsOrientationStatistics1972}, which shows that the MFD can be obtained by restricting the matrix Gaussian distribution to the Stiefel manifold. Consider a random matrix $Y\in \mathbbm{R}^{3\times 3}$ following the matrix Gaussian distribution with parameters $N\in \mathbbm{R}^{3\times3}$ and $0\prec \Sigma =\Sigma^T\in \mathbbm{R}^{3\times 3}$, and its PDF is given by
\begin{align}\label{eq:mGD}
    p(Y)&\propto {\rm exp}\left[-\frac{1}{2}\| Y-N \|_{\Sigma^{-1}}^2\right],
\end{align}
of which the negative log-likelihood measures the distance between $Y$ and $N$ using F-norm distance, and the weight $\Sigma^{-1}$ describes the anisotropy of the distribution. Confining (\ref{eq:mGD}) on $Y,N\in SO(3)$, then the PDF is rewritten as $ p(Y\in SO(3))\propto {\rm etr}\left( -\Sigma^{-1} + Y\Sigma^{-1}N^T \right)\propto {\rm etr}\left( \Sigma^{-1}N^TY \right)$, which is the same as the PDF of $Y\sim \mathcal{M}(\Sigma^{-1}N^T)$. Thus, the negative log-likelihood of MFDs can also be viewed as a distance between $Y$ and $N$. The distance is introduced based on the norm distance, instead of the geodesic distance underlying the intrinsic structure, of the embedded space.
\begin{remark}\label{re:no_pd}
    Since the parameter $F$ can be any matrix in $\mathbbm{R}^{3\times 3}$,  $\mathcal{M}(\Sigma^{-1}N^T)$ is a valid PDF of the MFD, even if $\Sigma^{-1}$ is not a positive definite matrix. More specifically, there is no need to choose a positive definite weighted matrix to make the normalization constant of the MFD finite.
\end{remark}

Similarly, BCDs are defined by restricting matrix Gaussian distributions on $\mathbbm{R}^{(3+n)\times (3+n)}$ to $SE_n(3)$.
\begin{definition}\label{de:BCD}
       The random extended transformation matrix $T=\mathcal{T}(R,X)\in SE_n(3)$ follows the BCD with parameters $M\in \mathbbm{R}^{3\times n}$, $U,V\in SO(3)$, $P\in \mathbbm{R}^{3\times n}$, $S={\rm diag}[s_1,s_2,s_3]$ and $K\in\mathbbm{S}^n_{++}$, if it has the following PDF as
    \begin{align}
     p(T) = \frac{1}{C} {\rm{etr}}\left\{-\frac{1}{2} {\left( {T - \bar T} \right)\left[ {\begin{array}{*{20}{c}}
\Lambda &P\\
{{P^T}}&K
\end{array}} \right]{{\left( {T - \bar T} \right)}^T}} \right\}\label{eq:de_BCD}
    \end{align}
    where $\Lambda = VSV^T+PK^{-1}P^T$ and $\bar T  = \mathcal{T}(UV^T,M) \in SE_n(3)$. Then, there is $T\sim \mathcal{B}(U,S,V,P,M,K)$.
\end{definition}
The normalized constant is calculated by integrating $p(T)$ with a Haar measure  $dT$. Following the Haar measure on $SE(3)$ \cite{oualkacha2012estimation}, $dT$ is defined by $dT=dRdX=dRd\overrightarrow{X}$, where $dR$ is the Haar measure of $SO(3)$ such that $\int_{SO(3)}dR=1$ and $dX=d\overrightarrow{X}=\Pi_{i=1}^n dx_i$ are the usual Lebesgue measures. According to Definition \ref{de:BCD}, a class of distributions on $SE_n(3)$ is defined as follows.
\begin{definition}\label{de:BCD_class}
    Let $\Theta \subseteq SO(3)\times \mathbbm{R}^3\times SO(3)\times\mathbbm{R}^{3\times n}\times\mathbbm{R}^{3\times n}\times \mathbbm{S}^n_{++}$ be a parameter set. A class of BCDs on $SE_n(3)$ is the set of probability distributions as
\begin{equation}
\mathfrak{B} = \left\{ \mathcal P_\theta|\theta=(U,S,V,P,M,K) \in \Theta \right\},
\end{equation}
where each $ \mathcal P_{\theta}$ has a density function defined by (\ref{eq:de_BCD}).
\end{definition}

The PDF given by (\ref{eq:de_BCD}) also indicates that the negative log-likelihood of BCDs, similarly to MFDs, can be viewed as the weighted F-norm distance between two elements of $SE_n(3)$ in the external space $\mathbbm{R}^{n\times n}$. As noted in Remark \ref{re:no_pd}, (\ref{eq:de_BCD}) also relaxes the requirement of positive definiteness for the attitude part of the weighted matrix.

For the convenience of integrating the PDFs, the marginal-conditional density functions of BCDs are presented. Subsequently, the normalization constant is derived.
\begin{theorem}\label{th:mc_pdf}
    Consider a random extended transformation matrix $T= \mathcal{T}(R,X)$ following $T\sim \mathcal{B}(U,S,V,P,M,K)$. Then the PDF of $T$ can be rewritten as
    \begin{align}\label{eq:MC_pdf}
        &p(T) =\frac{{{C_{MC}}}}{C}{\rm{etr}}\left[ -\frac{1}{2}{\left( {X - {M_C}} \right)K{{\left( {X - {M_C}} \right)}^T}} \right]\times \nonumber\\
        &{\rm{etr}}\left[ VSU^T R \right],
    \end{align}
    where ${C_{MC}} = {\rm{etr}}\left(-\frac{1}{2}VSV^T \right)$ and ${M_C} = M + (UV^T-R)P{K^{ - 1}}$.
\end{theorem}
\begin{proof}
    Let $\Lambda=VSV^T+PK^{-1}P^T$, and the PDF (\ref{eq:de_BCD}) can be rewritten as 
    \begin{align}
        &{\rm{etr}}\left\{-\frac{1}{2} {\left( {T - \bar T} \right)\left[ {\begin{array}{*{20}{c}}
\Lambda &P\\
{{P^T}}&K
\end{array}} \right]{{\left( {T - \bar T} \right)}^T}} \right\}\nonumber\\
&={\rm{etr}}\left[-\frac{1}{2} {\left( {R - UV^T
} \right)\Lambda {{\left( {R - UV^T
} \right)}^T} -} \right.\nonumber\\
&\left.\left( {X - M} \right){P^T}{{\left( {R - UV^T} \right)}^T}\! -\! \frac{1}{2}{\left( {X \!-\! M} \right)K{{\left( {X\! -\! M} \right)}^T}} \right]\label{eq:expand_BCD}\\
&= C_{MC}{\rm{etr}}\left[ -\frac{1}{2}{\left( {X - {M_C}} \right)K{{\left( {X - {M_C}} \right)}^T}} \right]{\rm{etr}}\left[ VSU^T R \right],\nonumber
    \end{align}
     where ${C_{MC}} = {\rm{etr}}\left(-S \right)$ and ${M_C} = M + (UV^T-R)P{K^{ - 1}}$. Thus, (\ref{eq:MC_pdf}) is verified.
\end{proof}
The theorem indicates that the marginal and conditional distributions are MFDs with $USV^T$, and matrix Gaussian distributions with $M_C$ and $K$, respectively. Using (\ref{eq:MC_pdf}), the normalized constant is calculated as follows.
\begin{proposition}
Suppose that $T\in SE_n(3)$ is characterized by $T\sim \mathcal{B}(U,S,V,P,M,K)$. Then the normalized constant of  $p(T)$ is given by $C = c_{\mathcal{M}}\left(S\right)(2\pi)^{({3}/{2})n}| K|^{-{3}/{2}}C_{MC}$.
\end{proposition}
\begin{proof}
    Intergrate $p(T)$ and there is
    \begin{align*}
&\int_{S{E_n}(3)} {p(T)} dT \\
&= \int_{SO(3)} {\int_{{^{3n}}} {\frac{{{C_{MC}}}}{C}{\rm{etr}}\left[ { - \frac{1}{2}\left( {X - {M_C}} \right)K{{\left( {X - {M_C}} \right)}^T}} \right]} }  \times \\
&{\rm{etr}}\left[ {VS{U^T}R} \right]dXdR\\
 &= \int_{SO(3)} {\frac{{{C_{MC}}{{(2\pi )}^{3/2n}}|K{|^{ - 3/2}}}}{C}} {\rm{etr}}\left[ {VS{U^T}R} \right]dR\\
& = \frac{{{C_{MC}}{{(2\pi )}^{(3/2)n}}|K{|^{ - 3/2}}{c_{\mathcal{M}}}\left( S \right)}}{C}.
    \end{align*}
    Since $\int_{S{E_n}(3)} {p(T)} dT =1$, the proposition is verified.
\end{proof}
This proposition shows that the relaxation of matrix positive definiteness does not affect the validity of $p(T)$.

\subsection{Conjugacy}
In this section, BCDs are proven to be conjugate for a class of widely used sampling distributions. The conjugacy of two distribution classes is defined formally as follows, with reference to \cite{gelman1995bayesian}.
\begin{definition}
    Supposing that $\mathfrak{F}$ is a class of  sampling distributions $ \mathcal P(Z|T)$, and $\mathfrak{P}$ is a class of prior distributions for $T$, then the class $\mathfrak{P}$ is conjugate for $\mathfrak{F}$ if the posterior
    \begin{align}
        \mathcal  P(T|Z)\in \mathfrak{P},\ \forall \mathcal  P(\cdot)\in \mathfrak{P},\  \mathcal P(\cdot|T)\in \mathfrak{F}.
    \end{align}
\end{definition}
Next, a useful transformation of the PDF of BCDs is presented as follows.
\begin{proposition}\label{pro:BCD_terms}
    For $T=\mathcal{T}(R,X)$ characterized by a BCD with parameters $\theta=(U,S,V,P,M,K)$, its PDF can be rewritten with linear and quadratic terms of $X$, a linear term of $R$, and a cross term involving both $R$ and $X$ as follows
    \begin{equation}
        p(T)\propto {\rm{etr}} \left[  C_R^1R^T +  XC_X^2 X^T +  XC_X^1 +  XC_R^XR^T \right ],\label{eq:BCD_terms}
    \end{equation}
    where $C_R^1 = USV^T+UV^TPK^{-1}P^T+MP^T$, $C_X^1 = P^TVU^T+KM^T$, $C_X^2 = -\frac{1}{2}K$ and $C_R^X =- P^T$. Parameters of the BCD can also be inverted from these coefficients by
    \begin{align}
&P = -(C_R^X)^T, \nonumber\\
&K = -2C_X^2, \nonumber\\
&(U,S,V) = {\rm pSVD}\!\left[C_R^1 - (C_X^1)^T K^{-1} P^T\right], \nonumber\\
&M = (C_X^1 - P^T V U^T)^TK^{-T}.\label{eq:BCD_coe}
\end{align}
\end{proposition}
\begin{proof}
    The PDF of $T$ can be expanded  as (\ref{eq:expand_BCD}) and then rewritten as
    \begin{align*}
        &p(T) \propto {\rm{etr}}\left[-\frac{1}{2} {\left( {R - UV^T
} \right)\Lambda {{\left( {R - UV^T
} \right)}^T} -} \right.\\
&\left.\left( {X - M} \right){P^T}{{\left( {R - UV^T} \right)}^T}\! -\! \frac{1}{2}{\left( {X \!-\! M} \right)K{{\left( {X\! -\! M} \right)}^T}}\right]\\
&\propto {\rm{etr}} \left[ \left( USV^T+UV^TPK^{-1}P^T+MP^T \right)R^T -\right .\\
&\left .\frac{1}{2}XKX^T +X(P^TVU^T+KM^T)-XP^TR^T \right],
    \end{align*}
    which verifies (\ref{eq:BCD_terms}). Substituting the expressions of the coefficients into (\ref{eq:BCD_coe}) term by term, the proposition is proved.
\end{proof}
The transformation in Proposition \ref{pro:BCD_terms} reveals that the log-probability density of BCDs consists of a linear combination of the terms $R$ and $X$. The proposition allows the fusion of BCDs and their conjugate distributions to be performed via simple algebraic addition of these coefficients, as shown below.

\begin{theorem}\label{the:BCD_con}
    Consider a class of sampling distribution on a p-dimensional manifold $\mathbbm{Y}$ as $\mathfrak{F}_\mathbbm{Y} = \{  \mathcal P(\tilde y|T)|\tilde y\in \mathbbm{Y}, T\in SE_n(3)\}$. The class of BCDs $\mathfrak{B}$ is conjugate for $\mathfrak{F}_\mathbbm{Y}$, if the PDF of $ \mathcal P(\tilde y|T)$ satisfies 
    \begin{align*}
       & p(\tilde y|T=\mathcal{T}(R,X))\propto {\rm etr} \left[ \tilde C_R^1(\tilde y){R^T} +X\tilde C_X^2(\tilde y){X^T} +\right .\\
        &\left . X\tilde C_X^1(\tilde y) + X\tilde C_R^X(\tilde y){R^T}+\tilde C_0(y)\right],\ \forall P(\tilde y|T)\in \mathfrak{F},
    \end{align*}
    where $\tilde C_R^1(\cdot):\mathbbm{Y}\rightarrow\mathbbm{R}^{3\times 3}$, $\tilde C_X^2(\cdot):\mathbbm{Y}\rightarrow\mathbbm{S}^{n\times n}_-$, $\tilde C_X^1(\cdot):\mathbbm{Y}\rightarrow\mathbbm{R}^{n\times 3}$, $\tilde C_R^X(\cdot):\mathbbm{Y}\rightarrow\mathbbm{R}^{n\times 3}$ and $\tilde C_0(\cdot):\mathbbm{Y}\rightarrow \mathbbm{R}^{3\times 3}$ are functions of $y$. If the prior distribution $ \mathcal P(T)\in \mathfrak{B}$ is generated by Proposition \ref{pro:BCD_terms} with coefficients $C_R^1$, $C_X^1$, $C_X^2$ and $C_R^X$, the coefficients of the posterior distribution $ \mathcal P(T|\tilde y)$ are given by $\hat C_R^1=C_R^1+\tilde C_R^1$, $\hat C_X^1=C_X^1+\tilde C_X^1$, $\hat C_X^2=C_X^2+\tilde C_X^2$ and $\hat C_R^X = C_R^X+ \tilde C_R^X$.
\end{theorem}
\begin{proof}
    Consider any $ \mathcal P(T)\in \mathfrak{B}$ with parameters $\theta=(U,S,V,P,M,K)$. We abbreviate $\tilde C(\tilde y)$ as $\tilde C$ for brevity. Using the Bayes' rule, the PDF of the posterior distribution $ \mathcal P(T|\tilde y)$ with a fixed sample $\tilde y$ is given by
    \begin{align*}
        &p(T|\tilde y) \propto p(T)p(\tilde y|T)\\
        &\propto {\rm{etr}} \left[ \left( C_R^1+\tilde C_R^1 \right)R^T + X(C_X^2+\tilde C_X^2) X^T + \right.\\
        &\left .   X(C_X^1+\tilde C_X^1) + X(C_R^X+\tilde C_R^X)R^T  \right ],
    \end{align*}
    where the terms $C_{(\cdot)}^{(\cdot)}$ are coefficients of the prior distributions given by Proposition \ref{pro:BCD_terms}. Let $\hat P=-(C_R^X+\tilde C_R^X)^T$, $\hat K=-2(C_X^2+\tilde C_X^2)$, $(\hat U,\hat S,\hat V) = {\rm pSVD}[C_R^1+\tilde C_R^1 -(C_X^1+\tilde C_X^1)^T\hat K^{-1}\hat P^T]$ and $\hat M=(C_X^1+\tilde C_X^1-\hat P^T\hat V\hat U^T)\hat K^{-T}$. Since $\hat{ K} = -2(C_X^2+\tilde C_X^2)= K+\tilde K$ is a positive definite matrix, and $\hat{U},\hat{S}$ as well as $ \hat{V}$ are given by pSVD, $(\hat U,\hat S,\hat V,\hat P,\hat M,\hat K)$ is a set of parameters that are valid under Definition \ref{de:BCD}. Thus, $p(T|\tilde y)\in \mathfrak{B}$ and the theorem is verified.
\end{proof}
Next, we consider two subsets of the sampling distribution class $\mathfrak{F}$, which are generated by direct state measurements and vector measurements, respectively. These two measurements are important for pose estimation in practice.
\begin{proposition}\label{pro:bayes_BCD_pose}
     Consider $T=\mathcal{T}(R,X),\tilde{T}=\mathcal{T}(\tilde R,\tilde X)\in SE_n(3)$. If the sampling distribution set ${\mathfrak{B}}_T$ is the subset of the BCD class $\mathfrak{B}$ satisfying
     \begin{equation}
         \mathfrak{B}_T = \{  \mathcal P_{\tilde{\theta}}(\tilde T|T)|\tilde{\theta}=(\tilde U,\tilde S,\tilde V,\tilde P,\tilde M,\tilde K),  \tilde U\tilde V^T = R, \tilde{M}=X\},\label{eq:Bt}
     \end{equation}
then $\mathfrak{B}_T$ is a subset of $\mathfrak{F}_{SE_n(3)}$ defined in Theorem \ref{the:BCD_con}.
\end{proposition}
\begin{proof}
    The equation (\ref{eq:Bt}) indicates that the PDF of $ \mathcal P(\tilde T|T)$ is
    \begin{align*}
        p(\tilde T|T) = \frac{1}{C} {\rm{etr}}\left\{-\frac{1}{2} {\left( {\tilde T - T} \right)\left[ {\begin{array}{*{20}{c}}
\tilde\Lambda &\tilde P\\
{{\tilde P^T}}&\tilde K
\end{array}} \right]{{\left( {\tilde T - T} \right)}^T}} \right\},
    \end{align*}
    where $\tilde \Lambda=\tilde  V\tilde S\tilde V^T+\tilde P\tilde K^{-1}\tilde P^T$. Thus, the PDF has the same form as a BCD with parameters $\tilde{\tilde \theta} = (\tilde R\tilde V,\tilde S,\tilde V,\tilde P,\tilde X,\tilde K)$.  By Proposition \ref{pro:BCD_terms}, $p(\tilde T|T)$ can be rewritten as
\begin{align*}
    &p(\tilde T|T) \propto\\
        &{\rm etr} \left[ {\tilde C_R^1(\tilde T){R^T} + X\tilde C_X^2(\tilde T){X^T} + X \tilde C_X^1(\tilde T)+ X\tilde C_R^X(\tilde T){R^T}} \right],
\end{align*}
    where $\tilde C(\cdot)$ are coefficients corresponding to $\tilde{\tilde \theta}$. According to Theorem \ref{the:BCD_con} for $\mathfrak{F}_{\mathbbm{Y} = SE_n(3)}$, the proposition is verified.
\end{proof}

\begin{proposition}\label{pro:bayes_BCD_landmark}
    Consider $T=\mathcal{T}(R,X)$ and $\tilde{y}\in \mathbbm{R}^3$. If the sampling distribution set $\mathfrak{L}$ of $ \mathcal P(\tilde y \mid T)$ is a subset of three-dimensional Gaussian distributions with mean $\mu \in \mathbbm{R}^3$ and $\Sigma \in \mathbbm{S}^3_{++}$ satisfying
\begin{align}
    \mathfrak{L} = \left\{  \mathcal P_{\tilde{\theta}}(\tilde y \mid T)\mid \tilde{\theta}=(\mu,\Sigma),\ \mu =\left( T^{-1}\tilde r\right)_{[1:3]},\ \Sigma = \sigma^2 I_3 \right\},
\end{align}
where $(\cdot)_{[1:3]}$ extracts the first three components of the vector, $\tilde r \in \mathbbm{R}^{3+n}$ is defined as $\tilde r = [r^T, j]^T$ with a reference vector $r\in \mathbbm{R}^3$ and an indicator vector $j\in \mathbbm{R}^n$ and $\sigma\in \mathbbm{R}$, then $\mathfrak{L}$ is the subset of $\mathfrak{F}_{\mathbbm{R}^3}$ defined in Theorem \ref{the:BCD_con}.
\end{proposition}
\begin{proof}
   Since $\mu =\left( T^{-1}\tilde r\right)_{[1:3]} =R^T(r-Xj) $, the PDF of $  \mathcal P_{\tilde{\theta}}(\tilde y \mid T)$ can be rewritten as
    \begin{align*}
&{p_{\tilde \theta }}(\tilde y|T) \propto {\rm{etr}}\left[ { - \frac{1}{{2{\sigma ^2}}}\left( {\tilde y - \mu } \right){{\left( {\tilde y - \mu } \right)}^T}} \right]\\
 &\propto {\rm{etr}}\left\{ { - \frac{1}{{2{\sigma ^2}}}\left[ {\tilde y{{\tilde y}^T} - 2\tilde y{r^T}R + 2Xj{{\tilde y}^T}{R^T} - } \right.} \right.\\
& {\left. {2r{j^T}{X^T} + Xj{j^T}{X^T}} \right]}\bigg \}.
    \end{align*}
    Let $\tilde C_R^1(\tilde y) = \sigma^{-2} r\tilde y^T$, $\tilde C_X^2(\tilde y)=-0.5\sigma^{-2}jj^T$, $\tilde C_X^1(\tilde y) = \sigma^{-2}jr^T$, $\tilde C_R^X (\tilde y)= -\sigma^{-2}j\tilde y^T$ and $\tilde C_0(\tilde y)=-0.5\sigma^{-2}\tilde y \tilde y^T$, and there is $\tilde C_X^2(\tilde y) \in \mathbbm{S}_{-}^{n\times n}$. Thus, $\mathfrak{L}\subset \mathfrak{F}_{\mathbbm{R}^3}$.
\end{proof}
\begin{remark}
    By selecting appropriate indicators and reference vectors, $\tilde y$ can correspond to different sensor measurement equations. Let the columns of $X=[x_1,x_2,\dots,x_n]$ represent position, velocity, and the landmarks to be estimated in $\mathcal{F}_I$. When $j=[1,0,0_{1\times(n-2)}^T]^T$ and $r\neq 0_{3\times 1}$, samples are characterized by $\tilde y = R^T(r-x_1)+v$ for $v\sim \mathcal{N}(0_{3\times 1}, \sigma^2 I_3)$, representing the measurement of a known landmark in $\mathcal{F}_B$. Moreover, when $[j]_1 =1$ and $[j]_k=-1$ for $2<k\leq n$, and $r= 0_{3\times 1}$, there is $\tilde y = R^T(x_k-x_1)+v$, representing the measurement of an unknown landmark in $\mathcal{F}_B$. The latter is more commonly associated with the SLAM problem. Moreover, the pure attitude measurements, such as magnetometer, can also be included in the proposition with $j=0_{n\times 1}$. Pure position measurements, such as GPS measurements in $\mathcal F_I$, are not covered by this proposition because they do not contain the $R$. However, they still belong to the general conjugate class in Theorem \ref{the:BCD_con}, since $\tilde y=Xj+v$ yields only quadratic and linear terms of $X$.
\end{remark}
When the sampling distributions belong to $\mathfrak{F}$ in Theorem \ref{the:BCD_con} and the prior is characterized by BCDs, the posterior distributions can be obtained through explicit analysis and are still members of the BCD class. For sampling distributions that do not belong to $\mathfrak{F}$, a two-stage method can be used. First, reconstruct the pose from measurements and approximate the underlying distribution of the pose using BCDs. Second, fuse it with the prior by Proposition \ref{pro:bayes_BCD_pose}. The optimal approximations of BCDs to underlying distributions are presented in the next section.

\subsection{Sufficient Statistics and Corresponding Moments}
In this subsection, the sufficient statistics of BCDs are established, which are proven to be finite-dimensional. Then, moments corresponding to the sufficient statistics are computed. The sufficient statistics of BCDs are constructed by Fisher-Neyman factorization theorem \cite{bernardo1994bayesian} and Proposition \ref{pro:BCD_terms} as follows.
\begin{proposition}
   Consider a set of mutually independent samples $\{T_i=\mathcal{T}(R_i,X_i)\}$ of $ \mathcal P_{\theta}(T)\in \mathfrak{B}$ for $i=1,2,\dots, N$, where  $\mathfrak{B}$ is the class of BCDs defined in Definition \ref{de:BCD_class} and $\theta=(U,S,V,P,M,K)$ are the distribution parameters. Then, a set of sufficient statistics for $ \mathcal P_\theta(T)$ is given by
   \begin{align*}
       {\mathcal{S}}(\{T_i\}) = \left( {\sum\limits_{i = 1}^N {{R_i},\sum\limits_{i = 1}^N {{X_i},\sum\limits_{i = 1}^N {X_i^T{X_i},\sum\limits_{i = 1}^N {R_i^T{X_i}} } } } } \right).
   \end{align*}
\end{proposition}
\begin{proof}
    Using Proposition \ref{pro:BCD_terms}, the PDF of $ \mathcal P_\theta (T)$ can be rewritten as
    \begin{align*}
        p_\theta(T)=f(\theta){\rm{etr}} \left[  C_R^1R^T +  XC_X^2 X^T +  XC_X^1 +  XC_R^XR^T \right ],
    \end{align*}
    where $C_{(\cdot)}^{(\cdot)}$ is the function of $\theta$. Thus, the joint distribution of $\{T_i\}$ is given by
    \begin{align*}
        &p_\theta \left( \{T_i\} \right) = \prod\limits_{i=1}^{N} {p_\theta (T_i)}\\
        &=\frac{1}{f(\theta)^N}{\rm{etr}}\left[ {{{\left( {C_R^1} \right)}^T}\sum\limits_{i = 1}^N R } \right]{\rm{etr}}\left( {C_X^2\sum\limits_{i = 1}^N {{X^T}X} } \right)\times\\
        &{\rm{etr}}\left( {C_X^1\sum\limits_{i = 1}^N X } \right){\rm{etr}}\left( {C_R^X\sum\limits_{i = 1}^N {{R^T}X} } \right).
    \end{align*}
    Let ${\mathcal{S}}({T_i}) = \left( {\sum\limits_{i = 1}^N {{R_i},\sum\limits_{i = 1}^N {{X_i},\sum\limits_{i = 1}^N {X_i^T{X_i},\sum\limits_{i = 1}^N {R_i^T{X_i}} } } } } \right)$, and there is $p_\theta( \{T_i\})=g(\theta,\mathcal S(\{T_i\}))h(\{T_i\})$, where $h(\{T_i\})\equiv1$ and is independent of $\theta$. According to the Fisher-Neyman factorization theorem \cite{bernardo1994bayesian}, ${\mathcal{S}}(\{T_i\})$ is a set of sufficient statistics for $ \mathcal P_\theta(T)$.
\end{proof}
Sufficient statistics contain all the information in the sample about the distribution parameters and thus, only moments corresponding to sufficient statistics are needed for parameter estimation. Moreover, $\mathcal{S}(\{T_i\})$ contains only the first-order term of $R$ due to the construction of BCDs, which allows estimators to avoid computing complex second- and even higher-order moments for parameter inference.

The moments of BCDs corresponding to ${\mathcal{S}}(\{T_i\})$ are given by the following theorem.
\begin{theorem}\label{th:moment}
    Suppose that $T=\mathcal{T}(R,X)\in SE_n(3)$ is characterized by $T\sim \mathcal{B}(U,S,V,P,M,K)$. Then,
    \begin{align}
        &E(R)=E_{\mathcal{M}(USV^T)}(R),\label{eq:E_R}\\
        &E(X)= M+\left[UV^T-E(R) \right]PK^{-1},\label{eq:E_X}\\
        &E(R^TX)\!=\!E(R)^TM\!+\![E(R)^TUV^T-I_3]PK^{-1},\label{eq:E_RX}\\
        &E(X^TX)=3K^{-1}+E(X)^TE(X)+K^{-1}P^T[I_3-\nonumber\\
        &E(R)^TE(R)]PK^{-1},\label{eq:E_XX}
    \end{align}
    where $E_{\mathcal{M}(USV^T)}(R)$ is the first moment of $R\sim \mathcal{M}(USV^T)$.
\end{theorem}
\begin{proof}
    $E(R)$ is directly obtained from the fact that the marginal distributions of BCDs are MFDs with $USV^T$. For $E(X)$, we integrate $X$ and get
    \begin{align*}
        &E(X) = \int_{SO(3)} {\int_{{\mathbbm{R}^{3 \times n}}} {Xp(X|R)p(R)dXdR} } \\
        & = \int_{SO(3)} {M + \left[ {U{V^T} - R} \right]} P{K^{ - 1}}dR\\
        &= M + \left[ {U{V^T} - E(R)} \right]P{K^{ - 1}}.
    \end{align*}
    Similarly, there is
    \begin{align*}
        &E(R^TX)  = \int_{SO(3)} {R^TM + \left[ {R^TU{V^T} - I} \right]} P{K^{ - 1}}dR\\
        &=  E{(R)^T}M + \left[ {E{{(R)}^T}U{V^T} - I} \right]P{K^{ - 1}}.
    \end{align*}
    Finally, $E(X^TX)$ is given by
    \begin{align*}
        &E(X^TX) = \int_{SO(3)} {\int_{{^{3 \times n}}} {{X^T}Xp(X|R)p(R)dXdR} }\\
        &=\int_{SO(3)} M_C^TM_C+3K^{-1} dR\\
        &=3K^{-1}+E(X)^TE(X)+K^{-1}P^T[I_3-E(R)^TE(R)]PK^{-1}.
    \end{align*}
\end{proof}
\section{Optimal Estimation} \label{sec:opt_est}
This section focuses on finding optimal approximations of the underlying distribution in the BCD class with moments or samples. 

\subsection{Mean Extended Transformation Matrix}\label{sec:mean}
Similarly to the first moment of MFDs on $SO(3)$, $E(T)$ does not necessarily belong to $SE_n(3)$, and a mean extended transformation matrix should be defined as an estimate of BCDs. The following theorem gives two means of BCDs by maximizing the probability and minimizing the mean square error, respectively.
\begin{theorem}\label{th:optimal_est}
    For a BCD denoted by $\mathcal{B}(U,S,V,P,M,K)$, the mean of the BCD that maximizes probability is given by
    \begin{equation}
        \hat T_{max}=\operatorname*{arg\,max}_{T\in SE_n(3)}\{ p(T) \}=\mathcal{T}(UV^T,M),\label{eq:estimate_max}
    \end{equation}
    and that minimizes the mean square error is given by 
    \begin{align}
      & \hat T_{mse}\!=\!\operatorname*{arg\,min}_{T\in SE_n(3)}{\int_{SE_n(3)}\!\|T-\tilde T\|^2p(\tilde T)d\tilde T}\nonumber\\
      & =\mathcal{T}\left(UV^T,E(X)\right).\label{eq:estimate_mse}
    \end{align}
\end{theorem}
\begin{proof}
    For any $T\in SE_n(3)$, 
    \begin{align*}
        &p(T) =\frac{{{C_{MC}}}}{C}{\rm{etr}}\left[ -\frac{1}{2}{\left( {X - {M_C}} \right)K{{\left( {X - {M_C}} \right)}^T}} \right]\times \nonumber\\
        &{\rm{etr}}\left[ VSU^T R \right]\leq\frac{{{C_{MC}}}}{C}{\rm{etr}}\left[ VSU^T R \right]\leq \frac{{{C_{MC}}}}{C}{\rm etr}[S],
    \end{align*}
    where the first equality holds if $X=M_C$ and the second equality holds if $R=UV^T$. Substituting $R=UV^T$ into $X=M_C$, we have $X=M$. Thus, there is
    \begin{equation*}
        p\left( \mathcal{T}(UV^T,M) \right)\geq p(T),\ \forall T\in SE_n(3),
    \end{equation*}
    which verifies (\ref{eq:estimate_max}).
    Next, using the definition of the Frobenius norm,
    \begin{align*}
        \|T-\tilde T\|^2 ={\rm{tr}}\left( {2{I_3} - 2R{{\tilde R}^T} + X{X^T} - 2\tilde X{X^T} + \tilde X{{\tilde X}^T}} \right).
    \end{align*}
    Substituting the norm into (\ref{eq:estimate_mse}), we have
    \begin{align*}
        &\int_{S{E_n}(3)} {\|T - \tilde T\|{^2}p(\tilde T)} d\tilde T\\
        &= 6 - 2{\rm{tr}}\left[ {RE{{\left( {\tilde R} \right)}^T}} \right] + {\rm{tr}}\left[ {X{X^T} - 2E\left( {\tilde X} \right){X^T}} \right] +\\
         &E\left( {\tilde X{{\tilde X}^T}} \right).
    \end{align*}
    Since $E\left( {\tilde X{{\tilde X}^T}} \right)$ and the constant $6$ are independent to variables to be optimized, 
    \begin{align*}
        &\operatorname*{arg\,min}_{T\in SE_n(3)}{\int_{SE_n(3)}\!\|T-\tilde T\|^2p(\tilde T)d\tilde T} = \\
        &\operatorname*{arg\,max}_{T\in SE_n(3)} \ 2{\rm{tr}}\left[ {RE{{\left( {\tilde R} \right)}^T}} \right] + {\rm{tr}}\left[ { 2E\left( {\tilde X} \right){X^T}-X{X^T}} \right],
    \end{align*}
    where
    \begin{align*}
        &2{\rm{tr}}\left[ {RE{{\left( {\tilde R} \right)}^T}} \right] + {\rm{tr}}\left[ { 2E\left( {\tilde X} \right){X^T}-X{X^T}} \right]\\
        &=2{\rm{tr}}\left[ {RE{{\left( {\tilde R} \right)}^T}} \right] -\left\| X-E\left( \tilde X \right) \right\|^2+\left\|  E\left( \tilde X \right)\right\|^2\\
        &\leq 2{\rm tr}\left[ D \right]+\left\|  E\left( \tilde X \right)\right\|^2
    \end{align*}
    with $D=U^TE\left( \tilde R \right)V$ given by (\ref{eq:MF_E}). The last equality holds when $R = UV^T$ and $X = E(\tilde X)$. Thus, (\ref{eq:estimate_mse}) is verified.
\end{proof}
Generally,  $\hat T_{\max}$ and $\hat T_{\mathrm{mse}}$ are not equal to each other due to the geometric constraint of $SO(3)$. More specifically, the mean attitude subject to the constraint, $UV^T$, is not equal to the arithmetic mean $E(R)$, which leads to $UV^T - E(R) \neq 0$ in
$E(X) = M + [UV^T - E(R)] P K^{-1}$. According to \cite{lee2018bayesian}, $\partial \ln c(S) / \partial s_i$ approaches $1$ as $s_i$ increases, indicating that $\hat T_{\max}$ and $\hat T_{\mathrm{mse}}$ are close to each other when the dispersion of $R$ is highly concentrated around $UV^T$. The estimate $\hat T_{\max}$ can be used to obtain a maximum a posteriori estimate in Bayesian estimators with BCDs, while $\hat T_{\mathrm{mse}}$ prioritizes long-term accuracy over maximum probability. 

\subsection{Moment Matching}
Many practical problems, such as uncertainty propagation along kinematics or dynamics, require an approximation of the underlying distribution on $SE_n(3)$ with a specific probability model for subsequent estimation. Moment matching provides a solution by determining the parameters of a BCD such that its low-order moments coincide with those of the underlying distribution. Detailed solutions are presented as follows.
\begin{theorem}\label{th:mm}
    Suppose that a random matrix $T=\mathcal{T}(R,X)\in SE_n(3)$ follows a non-degenerate distribution, i.e., its probability distribution is not concentrated on a single element of $SE_n(3)$. Given moments of  $T$, denoted by $\mathcal{E}=(E(R),E(X),E(R^TX),E(X^TX))$, the BCD $\mathcal{B}(\hat U,\hat S,\hat V,\hat P,\hat M,\hat K)$ whose moments satisfy $\hat{\mathcal{E}} = (E_{\mathcal{B}}(R),E_{\mathcal{B}}(X),E_{\mathcal{B}}(R^TX),E_{\mathcal{B}}(X^TX))=\mathcal{E}$ is given by
    \begin{align}
        &\left[ \hat U,D ,\hat V\right] = {\rm pSVD}\left( E(R) \right),\label{eq:mm_UV}\\
        &\hat K\!=\!3\left[ E\left(X^TX\right)\!-\!E(X)^TE(X)-B^TA^{-1}B \right]^{-1},\label{eq:mm_K}\\
        &\hat P=A^{-1}B\hat K,\label{eq:mm_P}\\
        &\hat M=E(X)-\left[\hat U\hat V^T-E(R)\right]\hat P\hat K^{-1},\label{eq:mm_M}
    \end{align}
     where $A=I_3-E(R)^TE(R)$ and $B=E(R)^TE(X)-E(R^TX)$, and $\hat S$ is solved by
    \begin{equation}
        D={\rm{diag}}\left[ {\frac{{\partial \ln c(\hat S)}}{{\partial {\hat s_1}}},\frac{{\partial \ln c(\hat S)}}{{\partial {\hat s_2}}},\frac{{\partial \ln c(\hat S)}}{{\partial {\hat s_3}}}} \right].\label{eq:mm_S}
    \end{equation}
\end{theorem}
\begin{proof}
    Denote by $G= E\left(X^TX\right)-E(X)^TE(X)-B^TA^{-1}B$. We first prove that $A$ and $G$ are positive definite. Since the distribution of $T$ is non-degenerate,
    \begin{align*}
        &E\left[ {{{\left( {T - E(T)} \right)}^T}\left( {T - E(T)} \right)} \right]\\
&=\left[ {\begin{array}{*{20}{c}}
A&{ - B}\\
{ - {B^T}}&{E({X^T}X) - E{{(X)}^T}E(X)}
\end{array}} \right]\in \mathbbm{S}^{n+3}_{++},
    \end{align*}
whose Schur complement with respect to $A$ is $ E({X^T}X) - E{{(X)}^T}E(X)-B^TA^{-1}B$, and is equal to $G$. Because $E\left[ \left( {T - E(T)} \right)^T\left( {T - E(T)} \right) \right]$ is positive definite, $A$ and the Schur complement $G$ are both invertible.

Since $E(R)$ is equal to the arithmetic mean of the MFD with the parameter $USV^T$, $U$, $S$ and $V$ are solved directly from (\ref{eq:MF_E}) as given by (\ref{eq:mm_UV}) and (\ref{eq:mm_S}). Then, premultiplying \eqref{eq:E_X} by the transpose of \eqref{eq:E_R} and subtracting \eqref{eq:E_RX}, i.e., $\eqref{eq:E_R}^T\times\eqref{eq:E_X}-\eqref{eq:E_RX}$, gives $APK^{-1}=B$. Substituting $PK^{-1}=A^{-1}B$ to \eqref{eq:E_XX} yields $G= 3K^{-1},$ which verifies \eqref{eq:mm_K}. $P$ is solved by $PK^{-1}=A^{-1}B$ with $K$, and $M$ is obtained by substituting $U,V,P$ and $K$ into \eqref{eq:E_X}.
\end{proof}
\begin{remark}
    (Uniqueness of the solution) Solving $S$ from $D$ is equivalent to the first moment matching problem of $\mathcal{M}(S)$, as presented in (Theorem 2.2, Lee). The solution is unique since MFDs are exponential probability density models. When the proper singular values of $E(R)$, i.e., the diagonal elements of $D$, are distinct, $U$ and $V$ are uniquely determined by the pSVD of $E(R)$. Denoting $PK^{-1}$ by $Y$, \eqref{eq:E_X} and \eqref{eq:E_RX} give a linear system of equations with respect to $Y$ and $M$, whose solution is unique because $A$ and $G$ are positive definite. $Y$ and the given expectations uniquely determine $K$ by \eqref{eq:E_XX}, which also gives a unique $P$ by $P=YK$. Thus, the solution of moment matching is unique. When the proper singular values of $E(R)$ are not distinct, the pSVD of $E(R)$ can give additional solutions as $U'=UO$ and $V'=VO$, where $O\in SO(3)$ such that $OSO^T=S$. However, since $U'V'^T=UOO^TV^T=UV^T$ and $U'SV'^T=USV^T$, $U'$ and $V'$ do not affect the uniqueness of $M$, $P$, and $K$. Moreover, PDFs of BCDs, defined by \eqref{eq:de_BCD}, with $U',V'$ and with $U,V$ are identical for all $T\in SE(3)$ due to $VSV^T=V'SV'^T$. Therefore, we do not differentiate between solutions $(U,S,V,P,M,K)$ and $(U',S,V',P,M,K)$ in the subsequent context.
\end{remark}
In moment matching, all parameters are computed by simple algebraic operations with moments except for $S$. While $S$ must be obtained from \eqref{eq:MF_E} by numerical iteration, its computation and approximation have been carefully studied in the literature \cite{lee2018bayesian}. More importantly, no additional numerical procedures are introduced when estimating parameters from moments compared to the MFD on $SO(3)$, despite $SE_n(3)$ being a more complex manifold.

We further clarify the optimality of moment matching from the perspective of distribution approximation. The KL divergence from information theory is introduced as a measure of discrepancy between probability distributions on the manifold $SE_n(3)$, which is defined as
\begin{align}
    {D_{KL}}\left( { \mathcal P| \mathcal Q} \right) = \int_{SE_n(3)} p\left( T \right) \ln \frac{p\left( T \right)}{q\left( T \right)} \, dT,
\end{align}
where $ \mathcal P$ and $ \mathcal Q$ are distributions on $SE_n(3)$ with PDFs $p$ and $q$, respectively. The following proposition shows that the BCD obtained via moment matching minimizes the KL divergence relative to the underlying distribution among the class of BCDs.

\begin{proposition}\label{pro:KL}
Let $  \mathcal P$ be an arbitrary distribution on $SE_n(3)$ with moments\\ $\mathcal{E}=(E(R),E(X),E(R^TX),E(X^TX))$. Supposing that $\hat s_1$, $\hat s_2$ and $\hat s_3$ are distinct, the BCD $ \mathcal P_{\hat \theta}$ with $\hat{\theta} = (\hat U,\hat S,\hat V,\hat P,\hat M,\hat K)\in\Theta$ satisfying
    \begin{equation}
     \mathcal P_{\hat \theta} = \operatorname*{arg\,min}_{ \tilde P\in \mathfrak{B}}\ D_{KL}\left(\mathcal P| \mathcal{\tilde P}\right),\label{eq:KL_opt}
\end{equation}
    is determined by moment matching with $\hat{\mathcal{E}}=\mathcal{E}$ in Theorem \ref{th:mm}, where $\mathfrak{B}$ is the distribution class of BCDs as given in Definition \ref{de:BCD_class}.
\end{proposition}
\begin{proof}
See Appendix A.
\end{proof}

\subsection{Maximum Likelihood Estimation}\label{sec:mle}
We then consider the MLE problem to construct a BCD from its samples. Suppose there is a set of samples $\{T_i=\mathcal{T}(R_i,X_i)\}$ for $i=1,2\dots,N$ from a BCD $\mathcal{P}_{\theta}\in \mathfrak{B}$ with parameter $\theta = (U,S,V,P,M,K)$.  Using the transformation in Proposition \ref{pro:BCD_terms}, the joint log-likelihood function is defined as
\begin{align}
    &\mathcal{L}\left(\theta|\{T_i\}\right)= -N{\ln } f(\theta ) + {\rm{tr}}\left[ {C_R^1{{\left( {\sum\limits_{i = 1}^N R_i } \right)}^T}} \right]+\nonumber\\
    &{\rm{  tr}}\left( {C_X^2\sum\limits_{i = 1}^N {{X_i^T}X_i} } \right)+{\rm{tr}}\left( {C_X^1\sum\limits_{i = 1}^N X_i } \right)+\nonumber\\
    &{\rm{ tr}}\left( {C_R^X\sum\limits_{i = 1}^N {{R_i^T}X_i} } \right),
\end{align}
where 
\begin{align}
    &\ln f(\theta) = \ln c_{\mathcal{M}}(S)+\frac{3}{2n}\ln(2\pi)-\frac{3}{2}\ln| K|-n\ln 2+\nonumber\\
    &{\rm tr}\left[ PK^{-1}P^T+MP^TVU^T+\frac{1}{2}MKM^T \right].
\end{align}
  The $\theta$ that maximizes $\mathcal{L}\left(\theta|\{T_i\}\right)$ is obtained as follows.
\begin{theorem}\label{th:MLE}
    Let $\{T_i=\mathcal{T}(R_i,X_i)\}$ for $i=1,2\dots,N$ be independent and identically distributed samples from a BCD $\mathcal{P}_{\theta}\in \mathfrak{B}$ with $\theta = (U,S,V,P,M,K)$. Denote  by $\bar R=\frac{1}{N}{\sum\limits_{i = 1}^N R_i }$, $\overline {{X^T}X} =\frac{1}{N}\sum\limits_{i = 1}^N {{X_i^T}X_i}$, $\bar X=\frac{1}{N} \sum\limits_{i = 1}^N X_i $ and $\overline{R^TX}=\frac{1}{N}\sum\limits_{i = 1}^N {{R_i^T}X_i}$, respectively. Suppose the diagonal elements of $S$ are distinct. Then the MLE for $\{T_i\}$ is obtained as
    \begin{align}
        &\left[  U,\bar D , V\right] = {\rm pSVD}\left( \bar R\right),\label{eq:mle_UV}\\
        & K=3\left( \overline{X^TX}-\bar X^T\bar X-\bar B^T\bar A^{-1}\bar B \right)^{-1},\label{eq:mle_K}\\
        & P=\bar A^{-1}\bar B K,\label{eq:mle_P}\\
        & M=\bar X-\left( U V^T-\bar R\right) P K^{-1},\label{eq:mle_M}
    \end{align}
     where $\bar A=I_3-\bar R^T\bar R$ and $B=\bar R^T\bar X-\overline{R^TX}$, and $\hat S$ is solved by
    \begin{equation}
        \bar D={\rm{diag}}\left[ {\frac{{\partial \ln c( S)}}{{\partial { s_1}}},\frac{{\partial \ln c( S)}}{{\partial { s_2}}},\frac{{\partial \ln c( S)}}{{\partial { s_3}}}} \right].\label{eq:mle_S}
    \end{equation}
\end{theorem}
\begin{proof}
    We utilize the relationship between the KL divergence and the log-likelihood function to verify the theorem. Using \eqref{eq:entropy} in the proof of Proposition \ref{pro:KL}, it follows that
    \begin{align}
        \operatorname*{arg\,min}_{ \tilde P\in \mathfrak{B}}\ D_{KL}\left(\mathcal P|\tilde  {\mathcal P}\right) = \operatorname*{arg\,max}_{ \tilde P\in \mathfrak{B}}\ -H(\mathcal{P}|\mathcal{P}_\theta).
    \end{align}
    Substituting \eqref{eq:BCD_terms} into $-H(\mathcal{P}|\mathcal{P}_\theta)$, there is
    \begin{align*}
        &-H(\mathcal{P}|\mathcal{P}_\theta)=
        \int_{SE_n(3)}p(T)\left\{ \ln \frac{1}{f(\theta)}+\right. \\
        & {\rm tr}\left( C_R^1R^T \right)+{\rm tr}\left( C_X^2X^TX \right) +{\rm tr}\left( C_X^1X \right)+{\rm tr}\left( C_X^R R^TX \right)\bigg\}\\
       & =\ln \frac{1}{f(\theta)}+ {\rm tr}\left[ C_R^1E(R)^T \right]+ {\rm tr}\left[ C_X^2E(X^TX) \right] +\\
       &{\rm tr}\left( C_X^1E(X) \right)+{\rm tr}\left[C_X^R E(R^TX) \right],
        \end{align*}
        which is equivalent to replace arithmetic means with expectations in $ \frac{1}{N}\mathcal{L}\left(\theta|\{T_i\}\right)$. Therefore, the solution of $\delta \mathcal{L} = \nabla_\theta \mathcal{L} \delta \theta = 0$ is the same as that of $\delta H= 0$, and the solution of the latter is given by Theorem \ref{th:mm} when the diagonal elements of $S$ are distinct. By replacing expectations with arithmetic means in \eqref{eq:mm_UV}, \eqref{eq:mm_K}, \eqref{eq:mm_P}, \eqref{eq:mm_M} and \eqref{eq:mm_S}, $\theta$ such that $\nabla_\theta \mathcal{L} = 0$ is given by \eqref{eq:mle_UV}, \eqref{eq:mle_K}, \eqref{eq:mle_P}, \eqref{eq:mle_M} and \eqref{eq:mle_S}. This completes the proof.
\end{proof}
Computing the MLE of distributions on nonlinear parameter manifolds is not trivial. Instead of replacing the joint MLE by marginal and conditional MLEs in \cite{wang2021matrix}, or directly computing the MLE by numerical optimization in \cite{darling2016uncertainty}, which lacks convergence or global optimality guarantees, the MLEs $U,V,P,M,K$ of the BCD are solved analytically with arithmetic means. The MLE $S$ is solved via the same method as that for MFDs, which is equivalent to solving nonlinear equations on $\mathbbm{R}^3$.

\section{Bayesian estimator on $SE_2(3)$} \label{sec:estor}
Utilizing BCDs, a Bayesian estimator on $SE_2(3)$ is constructed, motivated by the simultaneous estimation of position, attitude, and linear velocity of a rigid body using a gyroscope, an accelerometer, and landmark measurements. Benefiting from the introduction of conjugacy, the BCD-based estimator can smoothly deal with measurements containing position-attitude coupling terms via closed-form updates. Moreover, the propagation with moment matching is implemented by an analytical method, and its optimality is guaranteed by Proposition \ref{pro:KL}.

\subsection{System Model}\label{sec:estimator_models}
Denote the rotation from $\mathcal{F}_B$ to $\mathcal{F}_I$ by $R$, the position and the velocity of $\mathcal{F}_B$ relative to $\mathcal{F}_I$, expressed in $\mathcal{F}_I$, by $p$ and $v$, respectively. Suppose the angular velocity and proper acceleration in $\mathcal{F}_B$, denoted by $\omega$ and $a$, are measured by an unbiased gyroscope and accelerometer, denoted by $\tilde w $ and $\tilde a$, respectively.
Then, the stochastic differential equations on $SE_2(3)$ read 
\begin{align*}
    &R_t^T dR_t=\left( \tilde{\omega}_t dt+H_\omega dW_\omega  \right)_\times ,\\
    &d{v}_t=\left(R_t\tilde{a}_t + g\right)dt+H_adW_a,\\
    &d{p}_t=v_tdt,
\end{align*}
where $W_\omega$ and $W_a$ denote two independent three-dimensional Wiener processes, $H_\omega$ and $H_a$ are the strength matrices, and $g$ is the local gravitational acceleration under the flat Earth assumption. The above system originates from the well-known equations in \cite{barrau2016invariant}. According to \cite{barrau2018stochastic}, the motion equations are discretized by a time sequence $\{t_0,t_1,\dots\}$ with a fixed time step $h=t_{k+1}-t_k\in \mathbbm{R}^1$ as 
\begin{subequations}\label{eq:d_me}
\begin{align}
   R_{k+1} &= R_k \exp\!\left[ h\left( \tilde{\omega}_k  \right)_\times +\left(  H_\omega \Delta W_\omega \right)_\times \right], \label{eq:d_me_R}\\
   v_{k+1} &= v_k + R_k\tilde a_k h+ H_a\Delta W_a + gh, \label{eq:d_me_v}\\
   p_{k+1} &=p_k+ v_k h, \label{eq:d_me_p}
\end{align}
\end{subequations}
where $\Delta W_\omega$ and  $\Delta W_a \in \mathbbm{R}^3$ are the stochastic increments of the Wiener processes over a time step $h$, characterized by $H_\omega \Delta W_\omega\sim \mathcal{N}(0_{3\times 1}, hH_\omega H_\omega^T)$ and $H_a \Delta W_a\sim \mathcal{N}(0_{3\times 1}, hH_a H_a^T)$, respectively. Symmetric matrice $H_\omega$ and $H_a$ are denoted by $G_\omega$ and $G_a$, respectively.

The observations of the relative positions of $m$ known landmarks in $\mathcal{F}_B$ are used for the fusion and are modeled as
\begin{align}\label{eq:mea_mod}
    \tilde y^i_k = R_k^T(r^i - p_k) + \eta_y^i, \ \text{for } i = 1, \dots, m,
\end{align}
where $r^i$ is the coordinate of the $i$-th landmark in $\mathcal{F}_I$, $\tilde y^i_k$ is the corresponding measurement in $\mathcal{F}_B$, and $\eta_y^i$ is the additive noise characterized by $\eta_y^i \sim \mathcal{N}(0_{3\times 1}, \sigma^i I_3)$. The measurements are assumed to be mutually independent.

In the derivation of motion estimation algorithms, BCDs are used to describe the uncertainty of the state
$T_k=\mathcal{T}\left(R_k,X_k=[p_k,v_k]\right) \in SE_2(3)$.
In the propagation step, the BCD characterizing $T_k$ is propagated along \eqref{eq:d_me} by moment matching in Theorem \ref{th:mm} to obtain the KL-optimal approximation of the propagated distribution within the BCD class. The BCD is fused with the likelihood as the prior distribution to obtain the posterior distribution, which still belongs to the BCD class due to conjugacy. The estimate at $t_{k+1}$ is extracted from the posterior BCD by Theorem \ref{th:optimal_est}. The detailed methodology is presented as follows.

\subsection{Propagation}
The propagation of uncertainty is conducted by propagating moments in Theorem \ref{th:moment} and moment matching in Theorem \ref{th:mm}. The moments propagated along \eqref{eq:d_me} are first given by the following theorem. 
\begin{theorem}\label{th:e_p}
    Given the moments of $T_k=\mathcal{T}(R_k,X_k=[p_k,v_k])$ as $E(R_k)$, $E(X_k)$, $E(R_k^TX_k)$ and $E(X_k^TX_k)$, the moments of the propagated state $T_{k+1}$ are obtained as
    \begin{subequations}\label{eq:e_pro}
        \begin{align}
        &E(R_{k+1}) = E(R_k)\Lambda_\omega\exp \left[h(\tilde\omega_k )_\times\right]+\mathcal{O}(h^{1.5}),\label{eq:e_pro_R}\\
        &E({X_{k + 1}}) = \left[ {\begin{array}{*{20}{c}}
{E({p_{k + 1}})}&{E({v_{k + 1}})}
\end{array}} \right],\\
        &E(R_{k+1}^T X_{k+1})= \left[ {\begin{array}{*{20}{c}}
{E\left( {R_{k+1}^T{p_{k+1}}} \right)}&{E\left( {R_{k+1}^T{v_{k+1}}} \right)}
\end{array}} \right]\\
        &E\left( {{X_{k+1}}^T{X_{k+1}}} \right) = \left[ {\begin{array}{*{20}{c}}
{E\left( {p_{k+1}^T{p_{k+1}}} \right)}&{E\left( {p_{k+1}^T{v_{k+1}}} \right)}\\
{E{{\left( {p_{k+1}^T{v_{k+1}}} \right)}^T}}&{E\left( {v_{k+1}^T{v_{k+1}}} \right)}
\end{array}} \right]
    \end{align}
    \end{subequations}
    where $\Lambda_\omega =  I_3+\frac{h}{2} \left[  -{\rm tr}(G_\omega)I_3+G_\omega\right]$ and $\mathcal{O}(h^{n})$ includes at least the $n$th-order terms of $h$, and
    \begin{subequations}
        \begin{align}
        &E(p_{k+1}) = E(p_k)+E(v_k)h,\label{eq:e_pro_p}\\
        &E(v_{k+1})=E(v_k) + E(R_k)\tilde a_kh+gh,\label{eq:e_pro_v}\\
        &E(R_{k+1}^Tp_{k+1}) = \exp \left[ { - h{{\left( {{{\tilde \omega }_k}} \right)}_ \times }} \right]\Lambda_\omega E\left( {{R_k}^T{p_k}} \right)+\nonumber\\
        &\ \ h\exp \left[ { - h{{\left( {{{\tilde \omega }_k}} \right)}_ \times }} \right]E\left( {R_k^T{v_k}} \right) + {\mathcal{O}}({h^{1.5}}),\label{eq:e_pro_Rp}\\
        &E(R_{k+1}^Tv_{k+1})=\exp \left[ { - h{{\left( {{{\tilde \omega }_k}} \right)}_ \times }} \right]\Lambda_\omega E\left( {{R_k}^T{v_k}} \right)+\nonumber\\
        &\ \ h\exp \left[ { - h{{\left( {{{\tilde \omega }_k}} \right)}_ \times }} \right]\left[ \tilde a_k+E(R_k)^Tg \right] + {\mathcal{O}}({h^{1.5}}),\label{eq:e_pro_Rv}\\
         &E({p_{k + 1}}^T{p_{k + 1}})=E(p_k^Tp_k)+2hE(p_k^Tv_k)+h^2E(v_k^Tv_k),\\
        &E(p_{k+1}^Tv_{k+1}) = E\left(p_k^Tv_k\right)+h\tilde a_k^TE\left(R_k^Tp_k\right)+\nonumber\\
        &\ \ hg^TE(p_k)+hE(v_k^Tv_k)+h^2\tilde a_k^T E\left(R_k^Tv_k\right)+h^2g^TE(v_k)\nonumber\\
        & E(v_{k+1}^Tv_{k+1} )= E(v_k^T v_k)+(g^2+\tilde a_k^2)h^2+h{\rm tr}(G_a)+\nonumber\\
        &\ \ 2h\tilde a_k^TE(R_k^Tv_k)\!+\!2hg^TE(v_k)\!+\!2h^2 \tilde a_k^T E(R_k)^Tg,
    \end{align}
    \end{subequations}
\end{theorem}
\begin{proof}
 See Appendix B.
\end{proof}
Theorem \ref{th:e_p} implies that the residual terms of moment propagation all contain higher-order terms of the time step, i.e., $\mathcal{O}(h^{1.5})$, and excludes residuals related solely to estimation errors. Utilizing the moments of $T_{k+1}$, which is assumed to be characterized by $\mathcal{B}(U_k,S_k,V_k,P_k,M_k,K_k)$, and Theorem \ref{th:mm}, a BCD can be obtained as an approximation of the propagated distribution, referred to as the prior BCD and denoted by $\mathcal{B}(U_{k+1},S_{k+1},V_{k+1},P_{k+1},M_{k+1},K_{k+1})$. The prior BCD, according to Proposition \ref{pro:KL} and Theorem \ref{th:MLE}, minimizes the KL divergence and coincides with the joint MLE. Moreover, the entire procedure is analytic except for solving $S$ from $E(R_{k+1})$. In comparison, similar Bayesian estimators, such as \cite{wang2021matrix} and \cite{darling2016uncertainty}, provide only the conditional and marginal MLE, or numerical solutions of the optimization problems on manifolds due to the difficulty of obtaining the joint MLE.

\subsection{Fusion and Estimate Extraction}
 The prior BCD in the last subsection is fused with the likelihood distribution provided by the landmark measurement model \eqref{eq:mea_mod}. Let the indicator vector $j$ in Proposition \ref{pro:bayes_BCD_landmark} be $[1,0]^T$. Then, the sampling distribution $\mathcal{P}(\tilde y|T)$ corresponds to the distribution of $\tilde y^i_k$. Denote the PDF of the prior BCD by $p(T_{k+1})$ and the parameters of the posterior distribution by superscript $+$. Using Bayes' rule and Proposition \ref{pro:bayes_BCD_landmark}, the posterior density conditioned on available landmark measurements $\mathcal{Y}_{k+1} = \{\tilde y^i_{k+1}\}$ can be obtained by
 \begin{align*}
    &p\left( {{T_{k + 1}}|{{\mathcal{Y}}_{k + 1}}} \right) \propto p\left( {{T_{k + 1}}} \right)\prod\limits_{i = 1}^m {p\left( {\tilde y_{k + 1}^i|{T_{k + 1}}} \right)} \\
    & \propto\! {\rm{etr}}\!\left\{\! {\left[ {C_R^1 \!+\! \sum\limits_{i = 1}^m\! {\tilde C_R^1\left( {\tilde y_k^i} \right)} } \right]{R^T}\! +\! X\!\left[\! {C_X^2 \!+ \!\sum\limits_{i = 1}^m\! {\tilde C_X^2\left( {\tilde y_k^i} \right)} }\! \right]\!{X^T} \!+\! } \right.\\
    &\left. {\left[ {C_X^1 + \sum\limits_{i = 1}^m {\tilde C_X^1\left( {\tilde y_k^i} \right)} } \right]X + X\left[ {C_R^X + \sum\limits_{i = 1}^m {\tilde C_R^X\left( {\tilde y_k^i} \right)} } \right]{R^T}} \right\},
 \end{align*}
where $C^{(\cdot)}_{(\cdot)}$ are the coefficients generated from parameters of the prior BCD and $\tilde C^{(\cdot)}_{(\cdot)}$ are generated from Proposition  \ref{pro:bayes_BCD_landmark} using the parameters $\sigma^i$, $j$ and the measurement $\tilde y_k^i$. Because BCDs are conjugate to the sampling distribution $p\left( \tilde y_{k+1}^i \mid T_{k+1} \right)$, the posterior distribution $p\left( T_{k+1} \mid \mathcal{Y}_{k+1} \right)$ is still a BCD, referred to as the posterior BCD, and the coefficients of the posterior BCD are the algebraic sum of the prior BCD and likelihood distribution coefficients, i.e.,
$C_{(\cdot)}^{(\cdot),+} = C_{(\cdot)}^{(\cdot)} + \tilde C_{(\cdot)}^{(\cdot)}$. The parameters of the posterior BCD are directly obtained by the transformation in Proposition \ref{pro:BCD_terms}. The estimates of $T_{k+1}$ can be extracted by Theorem \ref{th:optimal_est}, using $\hat T_{max}$ or $\hat{T}_{mse}$. The simulation presented in the next section demonstrates the difference between the two estimates.

Similarly to Bayesian estimators with MFDs on $SO(3)$, the correction of the BCD-based Bayesian estimator is implemented by analytic algebraic operations, despite the more complex geometric structures involved. No approximation is introduced to obtain the posterior distribution with known prior BCDs and conjugate sampling distributions. The conjugate distribution class in Theorem \ref{the:BCD_con} covers a broad range of measurement models, including direct pose, position, attitude, and landmark measurements. As a result, there is no need to design complex approximation methods to obtain the posterior distribution for BCD-based Bayesian estimators in most cases.

\section{Simulations}
\label{sec:Sim}
In this section, the TFG-IEKF in \cite{barrau2022geometry}, the UKF on Lie groups (abbreviated as UKF-LG) in \cite{brossard2017unscented}, the BCD-based Bayesian estimator with maximum probability estimates (abbreviated as BE-BCD), and the BCD-based Bayesian estimator with minimum mean square error estimates (abbreviated as BE-BCD-MSE) are applied to simulated UAV motion estimation with measurement models in Section \ref{sec:estimator_models}. The Bayesian estimator with matrix Fisher-Gaussian distributions is not included in the comparison, because the estimators in \cite{wang2021matrix} and \cite{wang2023ins} are not applicable to landmark measurements. Designing a new estimator with MFGDs compatible with landmark measurements would require introducing a novel approximation method to obtain the posterior MFGDs, which is beyond the scope of this work.

The gravity vector is set as $g=[0,0,9.81]^T \rm m/s^2$, and the actual trajectory of the UAV is generated by tracking the desired trajectory  
$p(t)_{\rm desired}=[10\sin(0.1t), 10\sin(0.2t), 2\cos(0.2t)-0.1t-10]^T \rm m$ using the controller in \cite{Gamagedara2019}. We consider that the motion of the UAV is estimated using a gyroscope, accelerometer, and three landmark measurements characterized by \eqref{eq:d_me_R}, \eqref{eq:d_me_v}, and \eqref{eq:mea_mod}, respectively. The strength matrices of the Wiener processes are set as $H_\omega = 10^{-4}k_m\  \rm rad/s^{1/2}$ and $H_a = 10^{-3}k_m\ \rm m/s^{3/2}$, where $k_m$ is a scale factor varying for different cases. The additive noise of the landmark measurements $\eta_y^i$ is set as $\sigma^i = 0.05k_m$. The measurement frequencies of the gyroscope and accelerometer are $f=100\ \rm Hz$, and that of the landmark measurements is $1\ \rm Hz$. The simulation period is $T_{\rm sim}=100\ \rm s$, and each case is run $M_{\rm sim}=50$ times on a 16-core AMD Ryzen 9 7945HX CPU with 32\,GB of RAM.

Initial states input into the estimators are generated by
\begin{align}
    \hat{T}_0 = \exp\left( \xi_0^\wedge \right)\hat T_0^-,\ \xi_0 \sim \mathcal{N}(0_{9\times 1}, \Sigma_0),\label{eq:ini_state}
\end{align}
where the attitude component of $\hat T_0^-$ is set as the real attitude rotated about its first axis by $\pi/2$, and the position and velocity components are shifted from the real initial values by $[2,0,0]^T \rm m$ and $[0.5,0,0]^T \rm m/s$, respectively. The covariance matrix is
\[
\Sigma_0 =
\operatorname{diag}(
\sigma^2_{0,R}I_3,
\sigma^2_{0,p}I_3,
\sigma^2_{0,v}I_3).
\] Initial covariance matrices of state components are characterized by $\sigma_{0,R}=30k_i$, $\sigma_{0,p}=0.3k_i$, and $\sigma_{0,v}=0.1k_i$, with a scale factor $0<k_i\in \mathbbm{R}$. To fairly compare the proposed estimator and the Kalman filters, the initial parameters of the BCDs are computed by MLE using samples from \eqref{eq:ini_state} in every case.

\subsection{Effects of the initial uncertainty}
\begin{figure}[htbp]
	\begin{center}
		\includegraphics[width=0.84\linewidth]{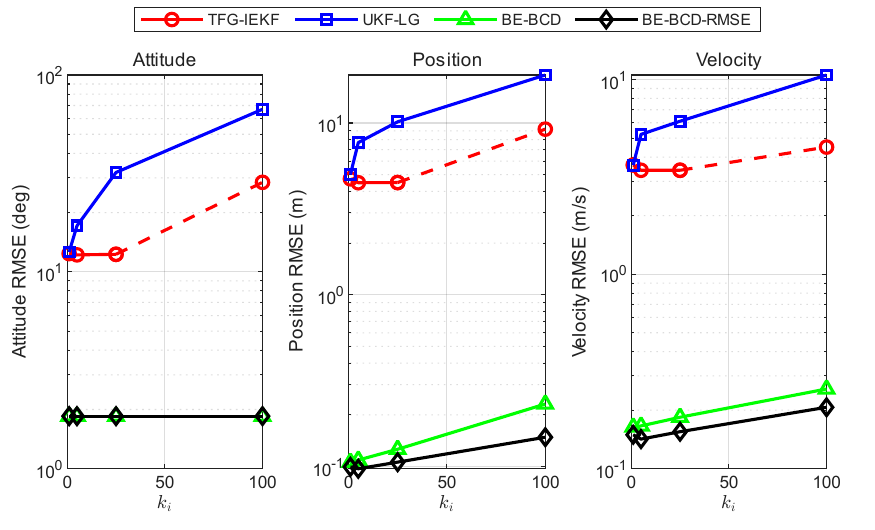}
		\caption{Effects of the initial uncertainty on RMSEs}\label{fig:E_ini_un}
	\end{center}
\end{figure}

\begin{figure}[htbp]
	\begin{center}
		\includegraphics[width=0.84\linewidth]{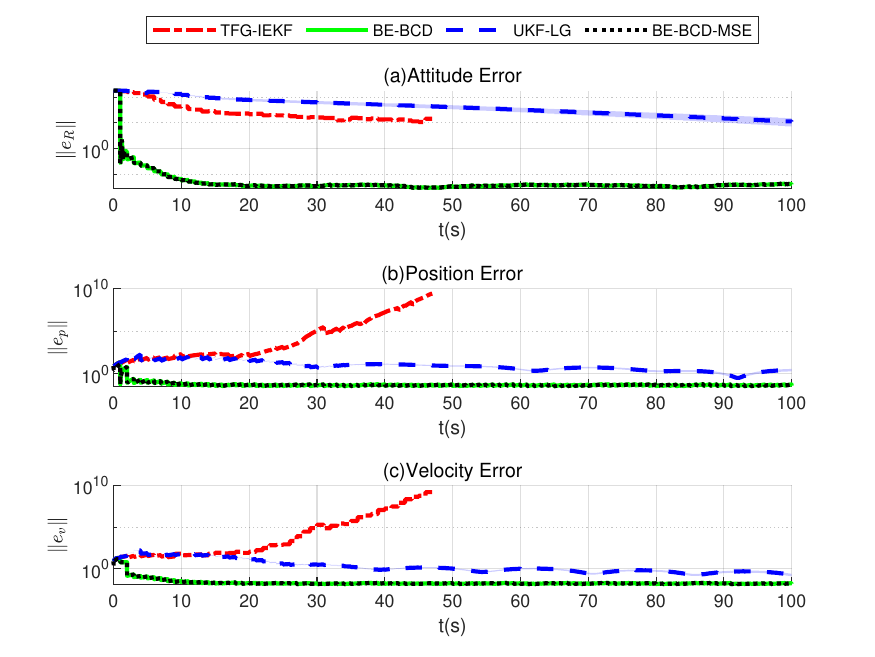}
		\caption{Estimation errors of estimators from 0s to 100s with $k_i=100$ and $k_m=1$}\label{fig:Ee_ini_un_100}
	\end{center}
\end{figure}

\begin{figure}[htbp]
	\begin{center}
		\includegraphics[width=0.84\linewidth]{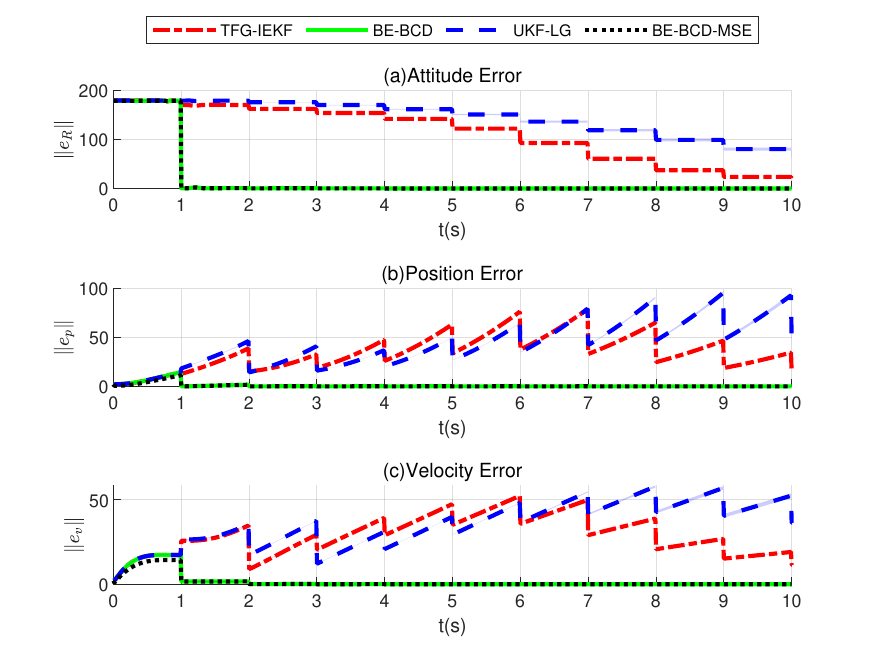}
		\caption{Estimation errors of estimators from 0s to 10s with $k_i=25$ and $k_m=1$}\label{fig:Ee_ini_un_25}
	\end{center}
\end{figure}

In this subsection, we investigate the effect of initial uncertainty on estimator performance. The scale factor $k_i$ is varied by $k_i \in \{1, 5, 25, 100\}$ to simulate different initial conditions, and $k_m$ is fixed to $1$. A larger $k_i$ indicates less confidence in \textit{a priori} information about the initial states, i.e., larger initial uncertainty. Moreover, because of the existence of initial bias, a small $k_i$ would also result in overconfidence in the input states. The performance of different state components of all estimators is evaluated using the Root-Mean-Square Error (RMSE) as
\begin{align*}
   & \| X \|_{RMSE} = \sqrt {\frac{1}{M_{sim}}\sum\limits_{j = 1}^{M_{sim}} {{{\left| {\frac{1}{{T_{sim}f}}\left( {\sum\limits_{k = 1}^{T_{sim}f} {\left\| e_{X,k}^j \right\|} } \right)} \right|}^2}} } 
\end{align*}
where variable $X$ is selected from $R$, $p$ and $v$, and $e_{X,k}^j$ are defined by $e_{R,k}^j={\ln(\hat R_{k|k}^j{R_k^T})^{\vee}}$, $e_{p,k}^j={\hat p_{k}^j- p_k}$ and $e_{v,k}^j={\hat v_{k}^j- v_k}$, respectively. The simulation results are summarized in Table \ref{tab:1} and Figs. \ref{fig:E_ini_un}, \ref{fig:Ee_ini_un_100} and \ref{fig:Ee_ini_un_25}.

\makeatletter
\long\def\@maketablecaption#1#2{%
  \@tablecaptionsize
  \global \@minipagefalse
  \hbox to \hsize{%
    \parbox[t]{\hsize}{\centering #1 \\ #2}%
  }%
}
\makeatother
\begin{table*}[htbp]
\centering
\caption{RMSE for simulation with different initial uncertainty*\tnote{*}}
\label{tab:1}


\begin{threeparttable}
{
\begin{tabular}{c|c|cccc}
\hline
$k_i$&             & TFG-IEKF& UKF-LG& BE-BCD& BE-BCD-RMSE\\ \hline
1& Att. RMSE(°)& 12.38(100\%)\tnote{**}& 12.611(102\%)& \textbf{1.85(15.0\%)}& \textbf{1.85(15.0\%)}\\ 
& Pos. RMSE (m)& 4.73(100\%)& 5.01(106\%)& \underline{0.106(2.23\%)}& \textbf{0.0994(2.10\%)}\\
 & Vel. RMSE (m/s)& 3.64(100\%)& 3.63(100\%)&\underline{ 0.1639(4.51\%)}&  \textbf{0.1495(4.11\%)}\\ \hline

5& Att. RMSE(°)
& 12.20(100\%)& 17.13(140\%)& \textbf{1.85(15.2\%)}& \textbf{1.85(15.2\%)}\\ 
& Pos. RMSE (m)
& 4.49(100\%)& 7.69(171\%)& \underline{0.109(2.43\%)}& \textbf{0.0970(2.16\%)}\\
 & Vel. RMSE (m/s)& 3.42(100\%)& 5.23(153\%)& \underline{0.166(4.86\%)}&\textbf{0.142(4.14\%)} \\ \hline

25& Att. RMSE(°)
& 12.28(100\%)& 31.98(260\%)& \textbf{1.85(15.1\%)}& \textbf{1.85(15.1\%)}\\ 
& Pos. RMSE (m)
& 4.50(100\%)& 10.16(226\%)& \underline{0.126(2.84\%)}&\textbf{0.106(2.35\%)}\\ 
& Vel. RMSE (m/s)& 3.42(100\%)& 6.11(179\%)& \underline{0.184(5.36\%)}& \textbf{0.155(4.52\%)}\\\hline

100& Att. RMSE(°)
& 28.5\tnote{***}& 66.98(100\%)& \textbf{1.85(2.76\%)}& \textbf{1.85(2.76\%)}\\ 
& Pos. RMSE (m)
& 9.21\tnote{***}& 19.03(100\%)& \underline{0.231(1.21\%)}& \textbf{0.148(0.778\%)}\\ 
& Vel. RMSE (m/s)& 4.50\tnote{***}& 10.56(100\%)& \underline{0.257(2.43\%)}& \textbf{0.207(1.96\%)}\\\hline
\end{tabular}
}

\begin{tablenotes}
    \footnotesize
    \item[*]  Best values are \textbf{bold}, and second best values are \underline{underlined}.
    \item[**] The percentage is computed with respect to the TFG-IEKF.
    \item[***] These RMSEs are obtained by excluding cases with singularity, and the percentage in this part is computed with respect to the UKF-LG.
\end{tablenotes}
\end{threeparttable}

\end{table*}
For $k_i \in \{1, 5, 25\}$, the attitude RMSEs of the two BCD-based estimators and the TFG-IEKF essentially remain unchanged, while the position and velocity RMSEs slightly increase. In contrast, those of the UKF-LG increase significantly. For $k_i = 100$, while the RMSEs of the two BCD-based estimators are still robust to initial uncertainty, the TFG-IEKF suffers numerical singularity for 2 out of 50 runs, as shown in Fig. \ref{fig:Ee_ini_un_100}. Even if the results of singular  runs are excluded, the RMSEs of the TFG-IEKF are still higher than those of the two BCD-based estimators. Thus, the proposed estimators exhibit better robustness to initial states in this set of simulations, i.e., the proposed estimators can more quickly filter out the negative impact of initial bias and prior confidence, which are always uncontrollable and difficult to difficult to calibrate in advance in practice. Fig. \ref{fig:Ee_ini_un_25} demonstrates that the first few fusions of the BCD-based estimators produce large and effective corrections, especially the first fusion at $t=1$, which results in lower RMSEs of these estimators. Effective correction of the proposed estimators mainly benefits from the fact that the structures of BCDs can retain the original measurements with their statistical characteristics, avoiding the approximation of Kalman filters in the correction step. These approximations are inaccurate because the assumption of small errors, especially for attitude components, is not well satisfied during the initial period when initial errors and uncertainties are large. Therefore, the attitude estimation errors of the two Kalman filters converge more slowly, and overshoots occur in their position and velocity errors before a marked reduction of attitude errors due to coupling of attitude and velocity in pose dynamics \eqref{eq:d_me}.

Moreover, the BE-BCD-MSE outperforms the BE-BCD in terms of position and velocity RMSEs, and the performance gap widens as the initial uncertainty increases. These results are consistent with the analysis in Section \ref{sec:mean}, which indicates that the two estimates are close when the attitude distribution is concentrated, and MSEs are more accurate in the long term. Figs. \ref{fig:Ee_ini_un_25} (b) and (c) also demonstrate that position and velocity errors of the BE-BCD-MSE are lower than those of BE-BCD before $t=1s$.

\subsection{Effects of the measurement noise}
\begin{figure}[htbp]
	\begin{center}
		\includegraphics[width=0.84\linewidth]{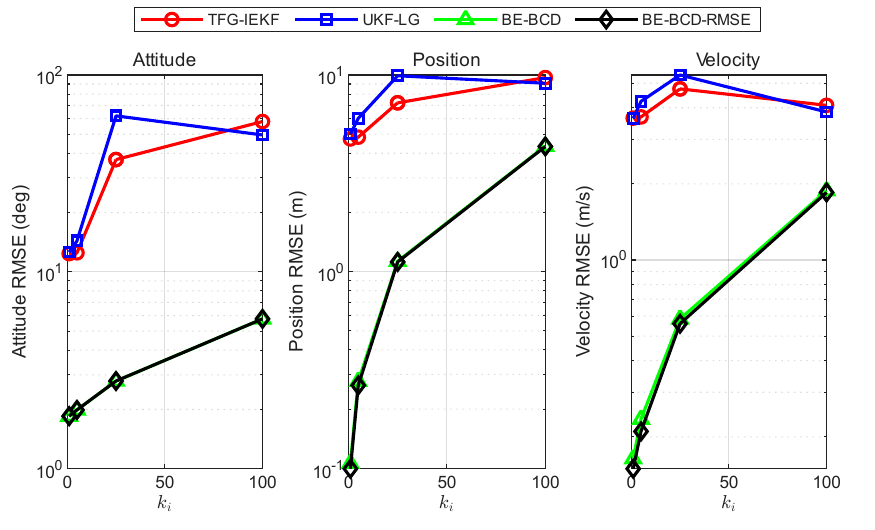}
		\caption{Effects of the measurement noise on RMESs}\label{fig:E_mea}
	\end{center}
\end{figure}

We investigate the effect of measurement noise on estimator performance in this subsection. The scale factor $k_m$ is varied by $k_m\in \{1, 5, 25, 100\}$ to simulate different measurement noise levels, while $k_i$ is fixed to $5$. A Larger $k_m$ may correspond to lower-cost sensors or poorer measurement conditions. The simulation results are summarized in Table \ref{tab:2} and Fig. \ref{fig:E_mea}.
\begin{table*}[htbp]
\centering
\caption{RMSE for different measurement noise*\tnote{*}}
\label{tab:2}


\begin{threeparttable}
{
\begin{tabular}{c|c|cccc}
\hline
$k_i$&             & TFG-IEKF& UKF-LG& BE-BCD& BE-BCD-MSE\\ \hline
1& Att. RMSE(°)& 12.38(100\%)\tnote{**}& 12.611(102\%)& \textbf{1.85(15.0\%)}& \textbf{1.85(15.0\%)}\\ 
& Pos. RMSE (m)& 4.73(100\%)& 5.01(106\%)& \underline{0.106(2.23\%)}& \textbf{0.0994(2.10\%)}\\
 & Vel. RMSE (m/s)& 3.64(100\%)& 3.63(100\%)&\underline{ 0.1639(4.51\%)}&  \textbf{0.1495(4.11\%)}\\ \hline

5& Att. RMSE(°)
& 12.5(100\%)& 14.36(115\%)& \textbf{1.99(15.9\%)}& \textbf{1.99(15.9\%)}\\ 
& Pos. RMSE (m)
& 4.83(100\%)& 6.02(124\%)& \underline{0.277(5.72\%)}& \textbf{0.264(5.47\%)}\\
 & Vel. RMSE (m/s)& 3.68(100\%)& 4.24(115\%)& \underline{0.235(6.40\%)}&\textbf{0.210(5.72\%)}\\ \hline

25& Att. RMSE(°)
& 37.19(100\%)& 62.0(166\%)& 2.79(7.51\%)& 2.79(7.51\%)\\ 
& Pos. RMSE (m)
& 7.22(100\%)& 9.91(137\%)& 1.13(15.7\%)&1.12(15.5\%)\\ 
& Vel. RMSE (m/s)& 4.75(100\%)& 5.39(113\%)& 0.585(12.3\%)& 0.560(11.8\%)\\\hline

100& Att. RMSE(°)
& 58.03(100\%)& 49.59(85.4\%)& \textbf{5.77(9.94\%)}& \textbf{5.77(9.94\%)}\\ 
& Pos. RMSE (m)
& 9.70(100\%)& 9.08(93.58\%)& \underline{4.35(44.85\%)}& \textbf{4.33(44.7\%)}\\ 
& Vel. RMSE (m/s)& 4.09(100\%)& 3.85(94.1\%)& \underline{1.88(46.0\%)}& \textbf{1.85(45.3\%)}\\\hline
\end{tabular}
}

\begin{tablenotes}
    \footnotesize
    \item[*]  Best values are \textbf{bold}, and second best values are \underline{underlined}.
    \item[**] The percentage is computed with respect to the TFG-IEKF.
\end{tablenotes}
\end{threeparttable}

\end{table*}

It is clear that the two proposed estimators are superior to the two Kalman filters in all four cases.  Although the position and velocity RMSEs of the BCD-based estimators show greater sensitivity to $k_m$ than the attitude RMSE, they remain lower than those of the TFG-IEKF and UKF-LG, even when $k_m=100$. For more practical noise measurement levels, such as $k_m\in\{1,5\}$, the promotion caused by the introduction of BCDs is remarkable. Additionally, the BE-BCD-MSE can still provide minorly better estimates than the BE-BCD.

\section{Conclusion}
This paper presents a novel class of Bayesian conjugate distributions on $SE_n(3)$ for modeling the uncertainty of motion states and constructing estimators. The distribution class is characterized by its conjugacy to widely used probability models of measurements in motion estimation, including direct pose measurements, position measurements, landmark measurements, and so on. The statistical properties of BCDs are investigated, particularly the optimal estimation in the BCD class. Two optimal means are derived as state estimates, and optimal parameter estimation from moments and sampling is presented. It is shown that these optimal estimates can be obtained analytically, avoiding local approximations or numerical iterations on manifolds in previous works. These properties improve the theoretical soundness and reliability of estimation with BCDs.  

A Bayesian estimator with BCDs is designed as an application to show how the distribution class is leveraged in motion estimation. Benefiting from the conjugacy and optimal estimation of BCDs, the estimator can give the minimum KL divergence approximation of the prior distribution in the BCD class and exact fusion with landmark measurements. Two optimal means are used as pose and velocity estimates, respectively. In numerical simulations, BCD-based estimators perform better in RMSE and transient response than the TFG-IEKF and UKF-LG with various initial uncertainty and measurement noise, and means with minimum squared error are more accurate than means with maximum probability as state estimates. Thus, it is believed that the proposed BCD class has potential applications in a wide range of estimation problems  involving uncertainty on $SE_n(3)$.

%

\bibliographystyle{unsrt}        
\bibliography{autosam}        

%

\appendix
\refstepcounter{section}
\section*{Appendix A: Proof of Proposition \ref{pro:KL}}
\addcontentsline{toc}{section}{Proof of Proposition}

\numberwithin{equation}{section}
Since 
    \begin{align*}
        {D_{KL}}( \mathcal P|\tilde  {\mathcal P}) = \int_{S{E_n}(3)} {p\left( T \right)\ln p(T) - p\left( T \right)} \ln \tilde  p(T)dT,
    \end{align*}
    and $ p\left( T \right)\ln  p(T)$ is independent of variables to be optimized, we have
    \begin{align}
        \operatorname*{arg\,min}_{ \tilde P\in \mathfrak{B}}\ D_{KL}\left(\mathcal P|\tilde  {\mathcal P}\right) = \operatorname*{arg\,min}_{ \tilde P\in \mathfrak{B}}\ \int_{S{E_n}(3)} -{ p\left( T \right)}  \ln \tilde p(T)dT.\label{eq:entropy}
    \end{align}
   Denote $\int_{S{E_n}(3)}- { p\left( T \right)} \ln \tilde p(T)dT$ by $H( \mathcal P|\tilde  {\mathcal P})$. Substituting the PDF  of $\mathcal{P}=\mathcal{B}(\hat U,\hat S,\hat V,\hat P,\hat M,\hat K)$ given in \eqref{eq:MC_pdf} into $H(\mathcal P|\tilde  {\mathcal P})$ leads to
    \begin{align*}
        &H( \mathcal P|\tilde  {\mathcal P}) =\int_{S{E_n}(3)}p(T)\left\{ \underbrace{\ln \frac{C}{C_{MC}}}_{p_1}+ \right.\\
        &\left. {\underbrace {{\rm{tr}}\left[ {\frac{1}{2}\left( {X - {M_C}} \right)K{{\left( {X - {M_C}} \right)}^T}} \right]}_{{p_2}} - \underbrace {{\rm{tr}}\bigg( {VS{U^T}R}\bigg )}_{{p_3}}} \right\}dT
    \end{align*}
Due to the geometric constraints of the parameters, we solve the optimization problem using a variational method. Since the variations of $U,V\in SO(3)$ can be written as $\delta U=U\omega^U_\times$, $\delta V=V\omega^V_\times$ for $\omega^V,\omega^U\in \mathbbm{R}^3$, respectively, and the variation of $K\in \mathbbm{S}_{++}^n$ satisfies $\delta K = \delta K^T$. For $p_1$, we have
\begin{align*}
    &\delta p_1 = \delta\left[  \ln c_{\mathcal{M}}\left(S\right)+\frac{3n}{2}\ln(2\pi)-\frac{3}{2}\ln| K|\right]\\
    &=\frac{\partial c_{\mathcal{M}}(s)}{\partial s}\delta s -\frac{3}{2}{\rm tr}\left( K^{-1} \delta K\right),
\end{align*}
where $s\in \mathbbm{R}^{3}$ satisfies $ {\rm diag}[s]=S$. Define an intermediate variable $A$ as $A=X-M_C$. The variation of $A$ is given by $\delta A = -\delta M_C$, where
\begin{align*}
    &\delta M_C = \delta M+ \left( \delta UV^T +U\delta V^T \right)PK^{-1}+\\
    &\left( UV^T -R \right)\delta P K^{-1}-\left( UV^T -R \right)PK^{-1}\delta KK^{-1}.
\end{align*}
Then
\begin{align*}
    &\delta p_2 = \frac{1}{2}{\rm tr}\left( \delta A KA^T+A\delta K A^T+AK\delta A^T \right)\\
    &={\rm tr}\left(- KA^T\delta M_C+\frac{1}{2}A^TA\delta K \right).\\
    &\delta p_3 = {\rm tr} \left( U^TRV{\rm diag}[\delta s] \right)+{\rm tr} \left( VS\delta U^TR \right)+\\
    &{\rm tr}\left( SU^TR\delta V \right).
\end{align*}
Substituting $\delta p_1$, $\delta p_2$ and $\delta p_3$ into $H(\mathcal P|\tilde  {\mathcal P})$, the variations of $H(\mathcal P|\tilde  {\mathcal P})$ with respect to $M$, $P$ and $S$ are given by
\begin{align*}
    &\delta _M H(\mathcal P|\tilde  {\mathcal P}) = \int_{SE(3)}-\tilde p(T){\rm tr }\left( KA^T\delta M \right)dT\\
    &= -{\rm tr}\left\{ KE(A)\delta M \right\},\\
    &\delta_P H (\mathcal P|\tilde  {\mathcal P})=\int_{SE(3)}-\tilde p(T){\rm tr }\left[KA^T\left(UV^T-R\right)\delta PK^{-1} \right]dT\\
    &=-{\rm tr}\left\{ \left[E(A)^TUV^T- E(A^TR) \right]\delta P\right\}\\
     &\delta_S H (\mathcal P|\tilde  {\mathcal P})=\int_{SE(3)}\tilde p(T)\left[ \frac{1}{c_{\mathcal{M}}(S)}\frac{\partial c_{\mathcal{M}}(S)}{\partial s}\delta s -\right.\\
     & {\rm tr} \left( U^TRV\delta S \right)\bigg]dT\\
     &=\frac{1}{c_{\mathcal{M}}(S)}\frac{\partial c_{\mathcal{M}}(S)}{\partial s}\delta s-{\rm tr} \left[U^TE(R)V\delta S \right],
\end{align*}
where $E(A)=\int_{SE_n(3)} p(T)(X-M_C)dT= E(X)-M-\left(UV^T-E(R)\right)PK^{-1}$ and 
\begin{align*}
    &E(A^TR)=\int_{SE_n(3)} p(T)ARdT\\
    &= E(R^TX)^T-M^TE(R)-K^{-1} P^T\left[ VU^TE(R)-I_3 \right].
\end{align*}
Using $\delta U=U\omega^U_\times$, $\delta V=V\omega^V_\times$ and $\delta K = \delta K^T$, the variations of $H(\mathcal P|\tilde  {\mathcal P})$ with respect to $U$, $V$ and $K$ are given by
\begin{align*}
    &\delta_U H(\mathcal P|\tilde  {\mathcal P}) \\
    &=\int_{SE_n(3)}p(T){\rm tr}\left[ -KA^T\delta UV^T PK^{-1} -VS\delta U^TR\right]dT\\
    &={\rm tr}\left[- E(A)^T\delta U V^TP +VS\omega_{\times}^UU^TE(R)\right],\\
    &\delta_V H(\mathcal P|\tilde  {\mathcal P})\\
    &=\int_{SE_n(3)}p(T){\rm tr}\left[ -KA^TU\delta V^T PK^{-1} -SU^TR\delta V\right]dT\\
    &={\rm tr}\left[ E(A)^TU \omega_{\times}^V V^TP -SU^TRV\omega_{\times}^V\right],\\
    &\delta_K H(\mathcal P|\tilde  {\mathcal P})\\
    &=\int_{SE_n(3)}p(T){\rm tr}\left\{-\frac{3}{2}K^{-1}\delta K +  \right.\\
    &\left . {\rm sym}\left[ A^T(UV^T-R)PK^{-1} \right]\delta K+\frac{1}{2}A^TA\delta K\right \}dT\\
    &=\int_{SE_n(3)}p(T){\rm tr}\left\{ -\frac{3}{2}K^{-1}\delta K+\left[ \frac{1}{2} (X-M)^T(X-M)-\right . \right .\\
    &\left. \left.\frac{1}{2}K^{-1}P^T(2I_3-VU^TR-R^TUV^T)PK^{-1}\right]\delta K \right\}\\
    & = {\rm{tr}}\left\{ {\left[ { - \frac{3}{2}{K^{ - 1}} + \frac{1}{2}E\left( {{X^T}X} \right) - {\rm{sym}}\left[ {E{{\left( X \right)}^T}M} \right] + } \right.} \right.\\
   &\left. {\left. {\frac{1}{2}{M^T}M - {K^{ - 1}}{P^T}\left( {{I_3} - {\rm{sym}}\left[ {V{U^T}E\left( R \right)} \right]} \right)P{K^{ - 1}}} \right]\delta K} \right\}
\end{align*}
where ${\rm sym}[\cdot]$ is defined by ${\rm sym}[B]=\frac{1}{2}(B+B^T)$ for $B\in \mathbbm{R}^{n\times n}$. 

By $\delta H(\mathcal P|\tilde  {\mathcal P})_{\tilde P=\hat{ P}} = 0$, $(\hat U,\hat S,\hat V,\hat P,\hat M,\hat K)$ is the solution of the set of equations $\delta _\varphi H(\mathcal P|\tilde  {\mathcal P})=0$ for $\varphi \in \{ U, S, V, P, M, K\}$. Since $K\in \mathbbm{S}^n_{++}$, $\delta_MH(\mathcal P|\tilde  {\mathcal P})=0$ is equivalent to
$ E(A) = E(X)-M-\left[UV^T-E(R)\right]PK^{-1}=0$, which implies
\begin{equation}
    E(X)=M+\left[UV^T-E(R)\right]PK^{-1}.\label{eq:EX_KL}
\end{equation}
Substituting $E(A)=0$ into $\delta_PH(\mathcal P|\tilde  {\mathcal P})=0$, $ \delta_P H (\mathcal P|\tilde  {\mathcal P})={\rm tr}\left[ E(A^TR) \delta P\right]=0$
and thus 
\begin{align}
    E(R^TX) = E(R)^TM+\left[E(R)^TUV^T-I_3 \right]PK^{-1}.\label{eq:RX_KL}
\end{align}
Rewriting $\delta_UH(\mathcal P|\tilde  {\mathcal P})=0$ and $\delta_VH(\mathcal P|\tilde  {\mathcal P})=0$ with $E(A)=0$, there are ${\rm tr}[U^TE(R)VS\omega_\times^U]=0$ and ${\rm tr}[SU^TE(R)V\omega_\times^V] = 0$. Since $\omega_\times^U$ and $\omega_\times^V$ are asymmetric matrices, this implies that $U^TE(R)VS$ and $SU^TE(R)V$ are symmetric, i.e., $DS=SD^T$ and $SD=D^TS$, where $D=U^TE(R)V$. Thus, $SDS^{-1}=D^T=S^{-1}DS$, i.e., $\frac{s_i}{s_j}D_{ij}=\frac{s_j}{s_i}D_{ij}$ for $i\neq j \in\{1,2,3\}$, implying $D_{ij}=0$ when $i\neq j$. Since $D$ is a diagonal matrix and $U,V \in SO(3)$, $\delta_SH(\mathcal P|\tilde  {\mathcal P})=0$ can be rewritten as
\begin{align}
    \frac{1}{c_{\mathcal{M}}(S)}\frac{\partial c_{\mathcal{M}}(S)}{\partial s_i} = D_{ii}, for \ i\in \{1,2,3\}.\label{eq:S_KL}
\end{align}
and $U$ and $V$ are given by
\begin{align}
    (U,D,V)={\rm pSVD}(E(R)).\label{eq:UV_KL}
\end{align}
Using \eqref{eq:EX_KL}, it follows
\begin{align*}
    &E{(X)^T}E(X) = \\
    &{\left[ {E(X) - M} \right]^T}\left[ {E(X) - M} \right] + 2{\rm{sym}}\left[ {E{{(X)}^T}M} \right] - {M^T}M\\
    &={K^{ - 1}}{P^T}\left\{ {I_3} - 2{\rm{sym}}\left[ {E{{(R)}^T}U{V^T}} \right] + \right.\\
    &\left. E{{(R)}^T}E(R) \right\}P{K^{ - 1}}+2{\rm{sym}}\left[ {E{{(X)}^T}M} \right] - {M^T}M,
\end{align*}
and thus, 
\begin{align*}
   & \delta_K H(\mathcal P|\tilde  {\mathcal P})= 
{\rm{tr}}\left\{ {\left[ {  \frac{1}{2}E\left( {{X^T}X} \right) - \frac{1}{2}E{{(X)}^T}E(X)-} \right.} \right. \\
&\left. {\left. {\frac{1}{2}{K^{ - 1}}{P^T}\left[ {{I_3} - E{{(R)}^T}E(R)} \right]P{K^{ - 1}}}- \frac{3}{2}{K^{ - 1}} \right]\delta K} \right\},
\end{align*}
which implies that $\delta_K H(\mathcal P|\tilde  {\mathcal P})=0$ is equivalent to
\begin{align}
    &E(X^TX)=3K^{-1}+E(X)^TE(X)+K^{-1}P^T[I_3-\nonumber\\
        &E(R)^TE(R)]PK^{-1}.\label{eq:XX_KL}
\end{align}
Equations \eqref{eq:EX_KL}, \eqref{eq:RX_KL} and \eqref{eq:XX_KL} are the same as \eqref{eq:E_X}, \eqref{eq:E_RX} and \eqref{eq:E_XX} in sequence. Combining \eqref{eq:S_KL} and \eqref{eq:UV_KL},
\begin{align}
    U{\rm diag}\left[ \frac{\partial\ln c_{\mathcal{M}}(S)}{\partial s_1}, \frac{\partial\ln c_{\mathcal{M}}(S)}{\partial s_2}, \frac{\partial\ln c_{\mathcal{M}}(S)}{\partial s_3}\right]V^T = E(R),
\end{align}
which is the same as \eqref{eq:E_R}. Therefore, the system of equations given by
$\delta H(\mathcal{P}\,|\,\tilde{\mathcal{P}})=0$ is the same as that given by $\hat{\mathcal{E}}=\mathcal{E}$, which indicates that the solution of \eqref{eq:KL_opt} coincides with the moment-matching BCD. This completes the proof.

\refstepcounter{section} 
\section*{Appendix B: Proof of Theorem \ref{th:e_p}}
\addcontentsline{toc}{section}{Proof of Theorem}

\setcounter{equation}{0}
\renewcommand{\theequation}{\thesection.\arabic{equation}}

 Since the  attitude kinematics are independent of other states, the propagation of $E(R_k)$ is consistent with Theorem 3.1 in \cite{leeBayesianAttitudeEstimation2018}, which verifies \eqref{eq:e_pro_R}. The propagation of $E(X_k)$ is directly obtained by taking the expectation of both sides of \eqref{eq:d_me_v} and \eqref{eq:d_me_p}. Denote by $\eta _\omega =H_\omega\Delta W_\omega$ and $\eta_a=H_a \Delta W_a$, respectively. Substituting \eqref{eq:d_me} into $R_{k+1}^Tp_{k+1}$ and $R_{k+1}^Tv_{k+1}$ yields
    \begin{align*}
       & R_{k+1}^Tp_{k+1} = \exp \left[ { - {{\left( h{{{\tilde \omega }_k} + {\eta _\omega }} \right)}_ \times }} \right]{R_k}^T\left( {{p_k} + {v_k}h} \right)\\
        &=\exp \left[ { - {{\left(h {{{\tilde \omega }_k}} \right)}_ \times }} \right]\exp \left[ { - {{\left( {{\eta _\omega }} \right)}_ \times } + {\mathcal{O}}({h^{1.5}})} \right]\left( {R_k}^T{p_k} +  \right.\\
        &\left. {R_k}^T{v_k}h\right)=\exp \left[ { -{{\left(  h{{{\tilde \omega }_k}} \right)}_ \times }} \right]\left[ {{I_3} - {{\left( {{\eta _\omega }} \right)}_ \times } + \frac{1}{2}{{\left( {{\eta _\omega }} \right)}_ \times }^2 + } \right.\\
        & {{\mathcal{O}}({h^{1.5}})} \bigg]\left( {{R_k}^T{p_k} + {R_k}^T{v_k}h} \right),\\
        &R_{k + 1}^T{v_{k + 1}} = \exp \left[ { - {{\left(h {{{\tilde \omega }_k} + {\eta _\omega }} \right)}_ \times }} \right] \left[ {{R_k}^T{v_k} + {{\tilde a}_k}h + {\eta _a}+ } \right.\\
        &\left .{R_k}^Tgh\right]=\exp \left[ { - {{\left(h {{{\tilde \omega }_k}} \right)}_ \times }} \right]\left[ {{I_3} - {{\left( {{\eta _\omega }} \right)}_ \times } + \frac{1}{2}{{\left( {{\eta _\omega }} \right)}_ \times }^2 + } \right.\\
        &\left. {{\mathcal{O}}({h^{1.5}})} \right]\left[ {{R_k}^T{v_k} + h{{\tilde a}_k} + {\eta _a} + {R_k}^Tgh} \right].
    \end{align*}
    Since $\eta_\omega$ and $\eta_a$ are mutually independent and independent of states, the expectations of $R_{k+1}^Tp_{k+1}$ and $R_{k+1}^Tv_{k+1}$ are given by
    \begin{align*}
        &E\left( R_{k+1}^Tp_{k+1} \right) = \\
        &\ \ \exp \left[ { - h{{\left( {{{\tilde \omega }_k}} \right)}_ \times }} \right]\left\{ I_3+\frac{h}{2} \left[  -{\rm tr}(G_k)I_3+G_k\right] \right\} E\left( {{R_k}^T{p_k}} \right)+\\
        &\ \ h\exp \left[ { - h{{\left( {{{\tilde \omega }_k}} \right)}_ \times }} \right]E\left( {R_k^T{v_k}} \right) + {\mathcal{O}}({h^{1.5}}),\\
        &E\left( R_{k+1}^Tv_{k+1} \right) = \\
         &\ \ \exp \left[ { - h{{\left( {{{\tilde \omega }_k}} \right)}_ \times }} \right]\left\{ I_3+\frac{h}{2} \left[  -{\rm tr}(G_k)I_3+G_k\right] \right\} E\left( {{R_k}^T{v_k}} \right)+\\
        &\ \ h\exp \left[ { - h{{\left( {{{\tilde \omega }_k}} \right)}_ \times }} \right]\left[ \tilde a_k+E(R_k)^Tg \right] + {\mathcal{O}}({h^{1.5}}),
    \end{align*}
    which verifies \eqref{eq:e_pro_Rp} and \eqref{eq:e_pro_Rv}.

    Substituting \eqref{eq:d_me_p} and \eqref{eq:d_me_v} into $p_{k+1}^Tp_{k+1}$, $p_{k+1}^Tv_{k+1}$, $v_{k+1}^Tv_{k+1}$, we have
    \begin{align*}
        &{p_{k + 1}}^T{p_{k + 1}} = {p_k}^T{p_k} + 2h{p_k}^T{v_k} + {h^2}v_k^T{v_k}\\
        &{p_{k + 1}}^T{v_{k + 1}} = p_k^T{v_k} + p_k^T{R_k}{{\tilde a}_k}h + {p_k}^T{\eta _a} + p_k^Tgh+\\
        &\ \ hv_k^Tv_k+h^2v_k^TR_k\tilde a_k+hv_k^T\eta_a+h^2v_k^Tg\\
        &v_{k+1}^Tv_{k+1} = v_k^T v_k+(g^2+\tilde a_k^2)h^2+\eta_a^T\eta_a+2v_k^TR_k\tilde a_kh+\\
        &\ \ 2v_k^T\eta_a+2v_k^Tgh+2h\tilde a_k^T R_k^T\eta_a+2h^2 \tilde a_k^T R_k^Tg+2h\eta_a^Tg .
    \end{align*}
    Then, it follows that
    \begin{align*}
        &E(p_{k + 1}^T{p_{k + 1}})=E(p_k^Tp_k)+2hE(p_k^Tv_k)+h^2E(v_k^Tv_k),\\
        &E(p_{k+1}^Tv_{k+1}) = E\left(p_k^Tv_k\right)+h\tilde a_k^TE\left(R_k^Tp_k\right)+hg^TE(p_k)+\\
        &\ \ hE(v_k^Tv_k)+h^2\tilde a_k^T E\left(R_k^Tv_k\right)+h^2g^TE(v_k)\\
        & E(v_{k+1}^Tv_{k+1} )= E(v_k^T v_k)+(g^2+\tilde a_k^2)h^2+h{\rm tr}(G_a)+\\
        &\ \ 2h\tilde a_k^TE(R_k^Tv_k)+2hg^TE(v_k)+2h^2 \tilde a_k^T E(R_k)^Tg,
    \end{align*}
    which completes the proof.

\end{document}